\documentclass[aps,prx,superscriptaddress,twocolumn,floatfix,showpacs,preprintnumbers,lengthcheck,longbibliography]{revtex4-2}

\usepackage[english]{babel}

\usepackage[utf8]{inputenc}
\usepackage{amsmath}
\usepackage{amssymb}
\usepackage{physics}

\usepackage{adjustbox} %
\usepackage{tikz}
\usepackage{mathtools}
\usetikzlibrary{quantikz}

\usepackage{newtxtext,newtxmath}	

\usepackage{bbold} %

\usepackage{xcolor}
\usepackage{graphicx}
\usepackage{hyperref}

\newcommand{\expv}[1]{\expval*{#1}}

\newcommand{\auckland}{\texttt{ibm\_auckland}}
\newcommand{\cairo}{\texttt{ibm\_cairo}}
\newcommand{\hanoi}{\texttt{ibm\_hanoi}} 
\newcommand{\mumbai}{\texttt{ibmq\_mumbai}}

\begin{document}

\title{Machine-learning certification of multipartite entanglement for noisy quantum hardware}

\author{Andreas J.~C.~Fuchs}
\thanks{These authors contributed equally to this work.}
\affiliation{Physikalisches Institut, Albert-Ludwigs-Universit\"{a}t Freiburg, Hermann-Herder-Stra{\ss}e 3, D-79104, Freiburg, Federal Republic of Germany}
\author{Eric Brunner}
\thanks{These authors contributed equally to this work.}
\affiliation{Quantinuum, Partnership House, Carlisle Place, London SW1P 1BX, United Kingdom}
\author{Jiheon Seong}
\affiliation{School of Electrical Engineering,
	Korea Advanced Institute of Science and Technology, Daejeon, Republic of (South) Korea}
\author{Hyeokjea Kwon}
\affiliation{School of Electrical Engineering,
	Korea Advanced Institute of Science and Technology, Daejeon, Republic of (South) Korea}
\author{Seungchan Seo}
\affiliation{School of Electrical Engineering,
	Korea Advanced Institute of Science and Technology, Daejeon, Republic of (South) Korea}
\author{Joonwoo Bae}
\affiliation{School of Electrical Engineering,
	Korea Advanced Institute of Science and Technology, Daejeon, Republic of (South) Korea}
\author{Andreas Buchleitner}
\affiliation{Physikalisches Institut, Albert-Ludwigs-Universit\"{a}t Freiburg, Hermann-Herder-Stra{\ss}e 3, D-79104, Freiburg, Federal Republic of Germany}
\affiliation{EUCOR center for Quantum Science and Quantum Computing, Albert-Ludwigs-Universit\"{a}t Freiburg, Hermann-Herder-Stra{\ss}e 3, D-79104, Freiburg, Federal Republic of Germany}
\author{Edoardo G.\ Carnio}
\affiliation{Physikalisches Institut, Albert-Ludwigs-Universit\"{a}t Freiburg, Hermann-Herder-Stra{\ss}e 3, D-79104, Freiburg, Federal Republic of Germany}
\affiliation{EUCOR center for Quantum Science and Quantum Computing, Albert-Ludwigs-Universit\"{a}t Freiburg, Hermann-Herder-Stra{\ss}e 3, D-79104, Freiburg, Federal Republic of Germany}


\begin{abstract}
Entanglement is a fundamental aspect of quantum physics, both conceptually and for its many applications. Classifying an arbitrary multipartite state as entangled or separable---a task referred to as the separability problem---poses a significant challenge, since a state can be entangled with respect to many different of its partitions.
We develop a certification pipeline that feeds the statistics of random local measurements into a non-linear dimensionality reduction algorithm, to determine with respect to which partitions a given quantum state is entangled. After training a model on randomly generated quantum states, entangled in different partitions and of varying purity, we verify the accuracy of its predictions on simulated test data, and finally apply it to states prepared on IBM quantum computing hardware.

\end{abstract}

\maketitle

\section{Introduction}
Entanglement is of fundamental importance in the field of quantum computation, since the faithful generation of highly entangled states is a necessary condition for any possible quantum speed-up~\cite{van2006universal, markov2008simulating, Nielsen_2012, zhou2020limits}.
Currently available quantum computing devices are noise- and error-prone~\cite{preskill2018quantum}, with diverse noise sources, like gate noise~\cite{he2020zeronoise, magesan2012efficient}, measurement noise~\cite{nachman2020unfolding}, systematic drifts~\cite{woitzik2023energy}, and cross-talk~\cite{sarovar2020detecting, seoMitigationCrosstalkErrors2021, Ketterer2023}.
It is therefore of great importance to understand which entangled states can be created on these devices. 
This is a fundamentally hard task, for two reasons.
First, the number of 
possible partitions of the state into sets of entangled parties
grows exponentially with
the number of subsystems---here qubits~\cite{coffman2000distributedentanglement}.
Second, for a mixed quantum state, it is difficult to discriminate classical from quantum correlations and, thus, to unambiguously determine the entanglement properties of the system \cite{Werner1989mixedstates,Uhlmann2000convexroof}.

Experimentally, obtaining the density matrix is already a challenge, since 
the number of measurements required for full state tomography scales exponentially with the number of qubits \cite{Nielsen_2012}, and is therefore only feasible for very small systems. 
Now that quantum registers with hundreds of qubits are available on current (noisy) quantum devices, we cannot rely on full state tomography to characterize their entanglement.
Alternative schemes exist where the number of measurements scales more favourably with the number of qubits. 
The multipartite concurrence, for instance, can be computed from just one measurement setting on two copies of the same given state \cite{mintert2005measures,aolita2006singleobservable}, although this method only yields a lower bound when dealing with mixed states
\cite{mintert2007observablemixed, Aolita2008scalable}.
Yet another possibility consists in witnessing the entanglement properties of a predefined target state: optimal schemes allow to calculate these witnesses from a number of measurements scaling linearly with the size of the register \cite{guhne2007toolbox}. 
Depending on the target state, however, these witnesses can be rather fragile against noise \cite{guhne2009entanglement}.
Unsurprisingly, methods to efficiently detect entanglement in larger systems---with as little  information as possible, for example from local random measurements~\cite{ketterer2019characterizing}---remain an active area of research \cite{cramer2010efficient, gross2010quantum, torlai2018neural, lu_entanglement_2018, koutny2023deeplearning, Ohnemus2023mpc}.

In this contribution we investigate how reliably we can prepare specific multipartite entangled states on current state-of-the-art quantum processors of the IBM Q System One~\cite{falcon} family. Due to the above-mentioned sensitivity of witnesses to noise, which makes the detection of entanglement ever harder for larger quantum registers, we compare their performance to a machine learning protocol, proposed in \cite{brunner_inprep,brunner_thesis} to identify entanglement partitions of mixed $N$-qubit quantum states. This scheme relies on a combination of random local measurements with a non-linear dimensionality reduction algorithm \cite{mcinnes2018umap}.
Since we train it with simulated random states, either pure or mixed, our model is by construction \emph{state agnostic}.
By giving up the detection of specific entangled states, our protocol proves much more resilient against hardware errors than the construction of suitable witnesses, at least for more than four qubits on the IBM hardware we tested.
Not only can we correctly resolve a large number of different entanglement partitions of the prepared states, we can also certify entanglement in states generated by increasingly complex quantum circuits, and thus unavoidably mixed by the hardware noise.
As a matter of fact, many mixed multipartite states, which we certify as entangled, would go otherwise undetected by the entanglement witnesses, due to the latter's inherent sensitivity to noise.

To find orientation in the manuscript: In Sec.~\ref{sec:state_preparation_on_quantum_hardware} we briefly describe state preparation on the aforementioned quantum hardware. In Sec.~\ref{sec:entanglementwitnesses} we present our experiments with the entanglement witnesses, while in Sec.~\ref{sec:results_entanglementpartitioncertification} and \ref{subsec:result_entanglementcertificationmixedstates} we construct and apply our statistical certification pipelines. In Sec.~\ref{sec:purity-estimation} we survey the possibility of estimating the purity of unknown states, as an extension of our approach. A summary of our insights and a further outlook is given in Sec.~\ref{sec:conclusions}. Many technical details, which are not immediately necessary to understand our results, are reported for reproducibility in Appendices \ref{appendix:witnesses}--\ref{app:all-snakes}.

\section{State Preparation on Quantum Hardware}
\label{sec:state_preparation_on_quantum_hardware}

Before we certify any entanglement, we need to define \emph{which} states we prepare, and \emph{how}.
In this paper we are interested in two states mainly---W and GHZ---that are well-studied in the field of entanglement theory~\cite{bengtsson2017geometry}.
The $N$-qubit W state~\cite{dur2000three} is defined as
\begin{align}
    \ket{\mathrm{W}_N} = \frac{1}{\sqrt{N}} (\ket{10...0} + \ket{01...0} + ... + \ket{0...01}),
\end{align}
while the $N$-qubit GHZ state~\cite{greenberger1990bell} is defined as:
\begin{align}
    \ket{\mathrm{GHZ}_N} = \frac{1}{\sqrt{2}} (\ket{0...0} + \ket{1...1}).
\end{align}

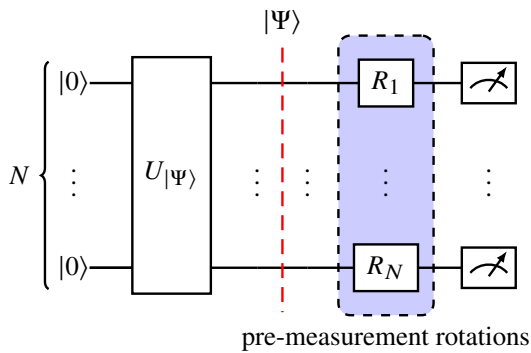
\begin{figure}[t]
\begin{adjustbox}{width=0.4\textwidth}
\begin{quantikz} \lstick[wires=3]{$N$}
	\ket{0}  &\gate[3, nwires=2]{U_{\ket{\Psi}}} & \qw \slice{$\ket{\Psi}$} & \qw & \gate{R_1} \gategroup[3,steps=1,style={dashed,
		rounded corners,fill=blue!20, inner xsep=2pt}, background, label style={label position=below,anchor=
		north,yshift=-0.2cm}]{{pre-measurement rotations}}  &  \meter{} \\
	\vdots  & &  \vdots & \vdots & \vdots & \vdots \\
	\ket{0}  & \qw &  \qw & \qw & \gate{R_{N}} & \meter{} 
\end{quantikz}
\end{adjustbox}
\caption{Illustration of a generic quantum circuit. The qubits are initialised in the state $\ket{0}^{\otimes N}$, then (up to the red dashed line) the state $\ket{\Psi}$ is prepared by the unitary $U_{\ket{\Psi}}$. Pre-measurement rotations $R_i$ (blue box) are applied right before the measurement in the computational basis, i.e., in the eigenbasis of the Pauli $Z$ operator. We prepare W and GHZ states on subsets of the register. The pre-measurement rotations are either random, or, for the entanglement witnesses, correspond to equidistant rotations about one of the axes of the Bloch sphere, see Sec.~\ref{sec:entanglementwitnesses} (together with App.~\ref{appendix:witnesses}) or~\cite{guhne2007toolbox}.}
\label{fig:circuit}
\end{figure}

Figure~\ref{fig:circuit} shows a generic quantum circuit that can be executed on currently available quantum hardware. 
In such a circuit, the qubit register is first initialised in the state $\ket{0}^{\otimes N}$, and then prepared in a target state $\ket{\Psi}$ by a unitary $U_{\ket{\Psi}}$. The latter is a specific, but not necessarily unique, sequence of single- and two-qubit gates, given, e.g., in~\cite{diker2016deterministic} for the W state, and in~\cite{woitzikEntanglementProductionConvergence2020} for the GHZ state, as well as in App.~\ref{appendix:GHZ_preparation}.
Since the measurement of the qubit's state projects it on either state $\ket{0}$ or $\ket{1}$ of the \emph{computational basis}, which is conventionally defined as the eigenbasis of 
the Pauli $Z$ ($\sigma_z$) operator, pre-measurement rotations are needed to measure the expectation value of other local operators $O_i$ on qubit $i$. 
By appropriately combining these rotations, we can measure the expectation value of any tensor product of local operators $ O_1 \otimes \dots \otimes O_N$.

At each step of the quantum circuit, i.e., during initialization, unitary evolution, and measurement, errors that are under extensive investigation \cite{de2021materials, kjaergaard2020superconducting, blatt2012quantum, bruzewicz2019trapped} can occur. 
These errors can be minimised by gate-sequence optimization, e.g., by choosing sequences that take the processor topology into account~\cite{nation2023suppressing}, or by post-processing the data, so called quantum error mitigation~\cite{cai2022quantum}.
Since we could not observe any significant difference in the results of our first simulations with and without error mitigation, we have decided against using this technique.

\subsection{Platforms \& Data Collection}
\label{subsec:platforms_datacollection}

We test our method on several IBM Quantum System One machines. 
We collected data in November 2022, May 2023 and June 2023, in prevalence on \texttt{ibmq\_ehningen} \footnote{This machine belongs to the `Falcon' family of variant r5.11 \cite{falcon}}, a machine with 27 qubits. 

In addition to the hardware, IBM also offers several \emph{simulators}, which allow the execution of quantum circuits with perfect initialization, readout and gate operations~\cite{qiskit2021}. On top of these simulations, IBM constructed \emph{noise models} that capture some of the sources of noise in currently available hardware. We will refer with \texttt{sim\_ehningen} to the (noisy) simulator based on IBM's noise model for \texttt{ibmq\_ehningen}.

We interact with the IBM computing platforms via submissions of runtime programs. %
These submissions we call a \textit{job}, and we collect time stamps of the calculations within each job.
A job consists of a list of circuits that are executed repeatedly to obtain the necessary statistics for the calculation of expectation values of local operators $ O_1 \otimes \dots \otimes O_N$.
The result of a single
 execution of a circuit like in Fig.~\ref{fig:circuit}, including the measurement, is referred to as a \textit{shot}.

\section{Entanglement Witnesses}
\label{sec:entanglementwitnesses}

Entanglement witnesses (EWs) are operators $\mathcal{W}$ which yield a non-negative expectation value for every (bi)separable state $\sigma$, i.e., $\Tr [\sigma \mathcal{W}] \geq 0$, and a negative expectation value for some witness-specific entangled states $\rho$~\cite{TERHAL2002313, RevModPhys.81.865, GUHNE20091, Chruscinski2014, Huber2019}. 
There exist several established approaches to construct EWs~\cite{PhysRevA.72.022340, PhysRevLett.94.060501, GUHNE20091, Huber2019}. 
Throughout our investigation, projector EWs, also called fidelity-based EWs~\cite{GUHNE20091}, serve our purpose (see Appendix~\ref{appendix:witnesses} %
for technical details of the implementation on the quantum hardware used for our analysis).  
The EW of an $N$-qubit W state reads \cite{GUHNE20091}:%
\begin{equation}
\label{eq:w_witness}
    \mathcal{W}_{\mathrm{W}_N} = \frac{N-1}{N} \mathbb{1} - \dyad{\mathrm{W}_N},
\end{equation}
with minimum value $-1/N$ attained by the W state itself. For an $N$-qubit GHZ state we have
\begin{equation}
\label{eq:ghz_witness}
    \mathcal{W}_{\mathrm{GHZ}_N} = \frac{1}{2} \mathbb{1} - \dyad{\mathrm{GHZ}_N},
\end{equation}
with minimum value $-1/2$ given by the pure GHZ state itself.
In general, entanglement witnesses only tolerate a finite amount of noise, beyond which they fail to detect entanglement.

To apply these EWs for entanglement detection on several IBM Quantum System One devices, we prepare entangled states on registers of $N=2,...,7$ qubits, using the circuit depicted in Fig.~\ref{fig:circuit}, with the appropriate unitaries, spelled out in Fig.~\ref{app:fig:ghz_circuits} below, for the preparation of GHZ and W states.
For one evaluation of either one of the entanglement witnesses in Eqs.~\eqref{eq:w_witness} or \eqref{eq:ghz_witness} on the thus prepared state, we use $10^5$ shots per expectation value of each of the local operators in Eqs.~\eqref{eq:M_k_GHZ} and \eqref{eq:M_k_W}.
Since noise affects each qubit independently (and dynamically \cite{woitzik2023energy}), this procedure -- preparation and witness evaluation -- is repeated for ten randomly chosen sets of connected qubits.

\begin{figure}
  \includegraphics[width=\columnwidth]{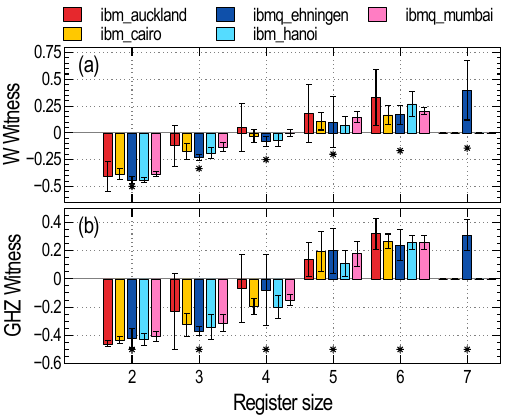}
\caption{Witness values $\langle\mathcal{W}\rangle$ (on the ordinates) of (a) W states and (b) GHZ states for several register sizes (on the abscissae) and devices (distinguished by colours). Values and error bars are, respectively, the arithmetic means and standard deviations of samples of ten values for each device.
Theoretical expectation values (marked by asterisks) are $-1/N$, with $N$ the register size, for the W witnesses, and $-1/2$ (independent of the register size) for the GHZ witnesses.
State-specific entanglement can be witnessed for states prepared on up to four qubits.
}
\label{fig:witness_scaling}
\end{figure}

We plot the resulting average witness values in Fig.~\ref{fig:witness_scaling}, where we observe that
either type of entanglement can be witnessed for up to four qubits on all processors, albeit with deviations from the theoretical values (asterisks) increasing with the number of qubits.
The exception here is \texttt{ibm\_auckland}, for which we fail to witness W-like entanglement already on some three-qubit registers, and which in general exhibits larger variances than the other quantum processors.
Beside these deviations, the inferred witness values are similar across the different processors.
We can therefore focus our analysis on \texttt{ibmq\_ehningen}~\footnote{It is worth mentioning that \texttt{ibm\_ehningen}, \texttt{ibm\_cairo}, \texttt{ibmq\_ehningen} and \texttt{ibm\_hanoi} are of the same processor type, i.e., Falcon r5.11, whereas \texttt{ibmq\_mumbai} is of type Falcon r5.10 \cite{pelofske2022quantumvolume}. In this latter reference `r5.10' is mislabeled as `r5.1'.}.

\begin{figure}
  \includegraphics[width=\columnwidth]{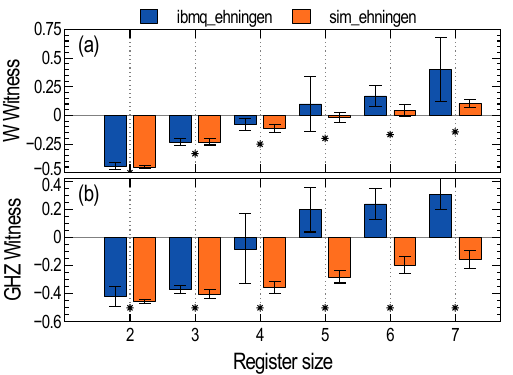}
	\caption{Witness values $\langle\mathcal{W}\rangle$ (on the ordinates) of (a) W states, and (b) GHZ states, for \texttt{ibmq\_ehningen} (blue) and its simulator \texttt{sim\_ehningen} (orange) for several register sizes (on the abscissae). 
 Thoretically expected witness values are marked by asterisks (see Fig.~\ref{fig:witness_scaling}). 
W states can be witnessed on registers of up to four qubits, on the real hardware and on the simulator.
 For GHZ states, entanglement can be witnessed for larger register sizes than four on the simulator, but not on the real hardware.
	}
	\label{fig:witness_noise_model}
\end{figure}

In Fig.\ \ref{fig:witness_noise_model}, we compare the witness values obtained with \texttt{ibmq\_ehningen} to those output by its (noisy) simulator \texttt{sim\_ehningen}. 
To the latter we feed the error rates (associated with qubit measurement and gates) returned by the real machine at the specific time of each witness estimation, since those error rates vary strongly over time \cite{woitzik2023energy}.
Not only do we observe large differences between the witness values of the device and the noisy simulator, the former are also more broadly spread than the latter. The cause of this deviation is hard to resolve, due to the complexity of the IBM hardware. 
Nevertheless, we know that current \cite{ibm_NoiseModel} device-extracted noise models include one- and two-qubit gate errors, arising from the action of a depolarizing channel and from thermal relaxation, and single-qubit readout errors on each measurement. 
There exist further sources of error that are not considered in the noise model, yet are known to significantly impact the fidelity of the state preparation \cite{Ketterer2023}, like cross-talk between the qubits and additional fluctuations~\cite{woitzik2023energy}, which may lead to the deviations observed in Fig.~\ref{fig:witness_noise_model}. 
Finally, depending on the preparation circuit employed, and its translation on the topology of the hardware, SWAP gates may become necessary to perform controlled gates between specific qubits (cf.\ Fig.~\ref{app:fig:ghz_circuits}).

In our analysis we see that the state-specific EWs do not detect W-type or GHZ-type entanglement for register sizes larger than four qubits.
The following scenarios may be playing out: 
1) the register was indeed prepared in an entangled state, though one which is sufficiently distinct from
the targeted GHZ or W state, such that it cannot be witnessed by the chosen witness, or
2) the noise in the state preparation is so strong that 
no entanglement can be imprinted onto
the register.
Recall that, when mixing an entangled state with the maximally mixed state \cite{Werner1989mixedstates},
there is a certain noise \emph{threshold} beyond which the resulting state is actually separable.
Indeed, 
all states within a finite distance from the maximally mixed state are separable \cite{zyczkowski_volume_1998,fine_equilibrium_2005,bengtsson2017geometry}, and we will hereafter encounter this threshold by following a continuous trajectory connecting
a given entangled state and a strongly mixed state.

In the next two sections we will explore both of the above scenarios by means of a statistical entanglement characterization protocol~\cite{brunner_inprep},
trained to recognize generic states from the entire Hilbert space (in contrast to witnesses constructed, as above, to target specific states).
With this state-agnostic tool, we are able to certify whether six qubits on IBM's hardware are prepared in a multipartite entangled (and otherwise arbitrary) state, and to faithfully discriminate cases with respect to which disjoint parts of the register are entangled.
By the same scheme, we can moreover analyze the progressive metamorphosis of entangled states to maximally mixed states, and ultimately identify 
a demarcation line between multipartite entangled and separable states. 
Thus we can determine whether the noise in IBM's hardware is too strong to prepare entangled states. 

\section{Classification of entanglement partitions}
\label{sec:results_entanglementpartitioncertification}

In a composite quantum system, like a qubit register, not all parties need to be entangled with each other, namely, entanglement can be witnessed only in some subsets of the register. The subsets in which we can split $N$ qubits are called partitions. 
For $N = 3$, for instance, the possible partitions are
\begin{align*}
	\mathbb{P} =  \big\lbrace & [[1,2,3]], \\
	&  [[3],[1,2]], [[1],[2,3]], [[2],[1,3]], \\
	& [[1],[2],[3]] \big\rbrace
\end{align*}
Each part $P_i$ of a partition $P = [P_1, \dots, P_r] \in \mathbb{P}$ describes a set of entangled qubits
(entangled \emph{within} rather than \emph{across} parts).
Two relevant examples for our later analysis of states in the entanglement partition $P = [P_1, \dots, P_r]$ are given by
\begin{equation}\label{eq:partitioned_GHZW_states}
\begin{split}
\ket{\mathrm{GHZ}_P} &= \ket{\mathrm{GHZ}_{P_1}} \otimes \dots \otimes \ket{\mathrm{GHZ}_{P_r}}, \\
\ket{\mathrm{W}_P} &= \ket{\mathrm{W}_{P_1}} \otimes \dots \otimes \ket{\mathrm{W}_{P_r}},
\end{split}
\end{equation}
of GHZ or W states on each part $P_i$, denoted with $\ket{\mathrm{GHZ}_{P_i}}$ and $\ket{\mathrm{W}_{P_i}}$, respectively.
In the case of single-qubit parts, we 
identify both $\ket{\mathrm{GHZ}_{P_i}}$ and $\ket{\mathrm{W}_{P_i}}$ with the balanced superposition of the computational basis states.

\subsubsection{Protocol}\label{sssec:protocol}
Our goal, as discussed in the previous section, is to establish whether the state prepared on the register of IBM hardware is entangled at all, even if not identically to the targeted W or GHZ state.
With the statistical protocol developed in \cite{brunner_inprep}, we can not only \emph{certify} entanglement of general mixed entangled states, but also discriminate the partition which describes the entanglement structure of the register state.

We consider an $N=6$ qubit register. 
Because the number of partitions grows fast with $N$, we restrict our attention to partitions ordered with respect to the number of constituent qubits of its parts, increasing from left to right. 
For the sake of brevity, we refer to these ordered partitions by
a sequence of integers, each given by the length of the ordered partition's parts,
e.g., 15 for partition [[1],[2,3,4,5,6]], and 1122 for partition [[1],[2],[3,4],[5,6]].

For each of 
the eleven ordered partitions of six qubits, indicated in Fig.~\ref{fig:entanglement_partition_assignment},
we generate 200 random pure states of the form
\begin{equation}\label{eq:ent_class:random_pure_part_state}
	\ket*{\psi} = \ket*{\psi_{P_1}} \otimes \dots \otimes \ket*{\psi_{P_r}} \,,
\end{equation}
with each $\ket{\psi_{P_i}}$ drawn from the Haar measure on the corresponding Hilbert space \footnote{This is done by generating a random Haar unitary in $2^{N_i}$ dimensions (with $N_i$ the number of qubits in part $P_i$) and applying it to an arbitrary but fixed $N_i$-qubit state.
For this we use the QuTiP \cite{johansson_qutip_2013} implementation of the algorithm proposed in \cite{mezzadri_how_2007}.}.
For each state in Eq.~\eqref{eq:ent_class:random_pure_part_state} we measure $k$-point correlators ($k=1, \dots, N$) of the many-qubit state in randomly chosen bases of each single qubit [cf.\ Eq.~\eqref{eq:ent_class:random_correlator}].
The correlators follow a distribution (determined by the random sampling of measurement directions) of which we calculate the second moment [cf.\ Eq.~\eqref{eq:ent_class:statistical_moments}], as proposed in \cite{ketterer2019characterizing,brunner_inprep} (similar in spirit to \cite{walschaers_statistical_2016}).
This dataset is easily accessible on any quantum device, since, provided the preparability of the underlying state, it only requires single-qubit rotations and measurements.
We split the data into $90\%$ training and $10\%$ test data.
With the training dataset we train a two-dimensional UMAP (Uniform Manifold Approximation and Projection---see App.~\ref{appendix:UMAP}) embedding model, which maps the high-dimensional data (containing the second moments of all $k$-point correlators, $k=1, \dots, N$) into the Euclidean plane $\mathbb{R}^2$ \cite{brunner_inprep} spanned, in Fig.~\ref{fig:entanglement_partition_assignment}, by the axes $x_0$ and $x_1$.
The thus obtained two-dimensional data representation yields a visualization, in the plane, of the various partitions sampled by the random $N$-qubit states in Eq.~\eqref{eq:ent_class:random_pure_part_state}.
This protocol, i.e., the generation of synthetic training data and the training of the UMAP model, is efficient---in the sense that the computational cost scales only polynomially in $N$---if we restrict the dataset to correlators of order $k \leq N_0$ with a constant $N_0 < N$, see \cite{brunner_inprep} and also App.~\ref{appendix:UMAP} for a short discussion.
In the following, however, we will, as a proof of concept, consider a comparatively small $(N=6)$-qubit system and include all correlation orders in the datatset.

The two-dimensional embedding of the training dataset is shown in Fig.~\ref{fig:entanglement_partition_assignment}, where we observe that well-separated clusters emerge for each partition (note that UMAP has no access to the partition underlying the scrutinized random states).
Based on this embedding, we train a decision-tree classifier, see App.~\ref{appendix:UMAP}, to consistently associate unseen data to one of the clusters visible in Fig.~\ref{fig:entanglement_partition_assignment}.
With this, we have built a classification pipeline that receives the second moments of the $k$-point correlators, and returns the label of a cluster (and, thus, the partition).
As a simple quantifier for the performance of the classification pipeline, we consider the accuracy score
\begin{equation}\label{eq:accuracy}
A = \frac{N_\mathrm{correct}}{N_\mathrm{correct} + N_\mathrm{incorrect}} \,,
\end{equation}
where $N_\mathrm{correct}$ and $N_\mathrm{incorrect}$ are the numbers of correct, respectively incorrect, classifications of points in a given dataset.
For the test dataset we obtain a high accuracy of $A = 0.98$ \footnote{Recall that the test dataset was generated in conjunction with, and therefore carries the same statistical properties as, the training dataset. The accuracy score for other datasets will be, in general, different.}.

\subsubsection{Classification of hardware-generated states}

\begin{figure*}
    \includegraphics[width=\textwidth]{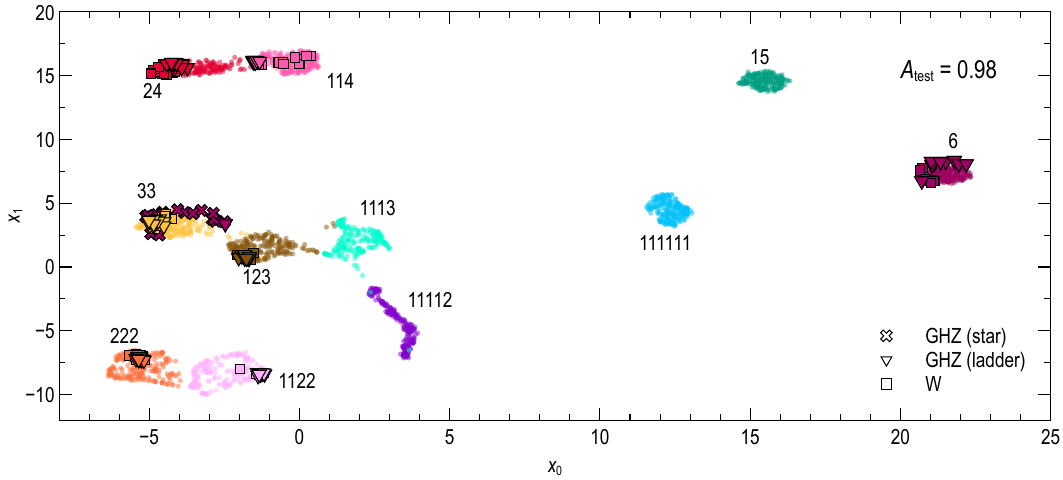}
\caption{
Assignment by the UMAP algorithm of W and GHZ states prepared on quantum hardware to the corresponding entanglement partitions, represented by coloured clouds of embedded training data points.
The $x_0$ and $x_1$ axes describe the two dimensions identified during training by the UMAP algorithm, and have no obvious physical interpretation.
Each cloud is obtained from 200 simulated random pure states, Eq.~\eqref{eq:ent_class:random_pure_part_state}, entangled according to the partition distinguished by the colour and identified by the adjacent label (cf.~Sec.~\ref{sssec:protocol} for the labelling scheme). 
Ten to fifteen quantum states, Eq.~\eqref{eq:partitioned_GHZW_states}, are prepared and measured on IBMQ Ehningen for each partition, and their embedding into the $(x_0,x_1)$ plane is represented by the larger symbols against the clouds of the training dataset: squares for W states, and triangles (resp.\ crosses) for GHZ states prepared with the `ladder' (resp.\ `star') circuit. The colour of these symbols indicates the \emph{intended} partition chosen upon preparation on the quantum hardware, such that a mismatch between a symbol's colour and the colour of its background cloud indicates that the preparation failed to generate the desired entanglement structure.
Given the high accuracy score ($A_\text{test} = 0.98$) on the test dataset, we conclude that we reliably prepared entangled states according to their intended entanglement  partition, except for the fully entangled GHZ states prepared with the `star' circuit (maroon crosses). In this case, all states  are assigned to partition 33 (cloud of yellow points). 
 } \label{fig:entanglement_partition_assignment}
\end{figure*}

We have now all the tools to certify the entanglement in a six-qubit register of \texttt{ibmq\_ehningen}.
For a given partition \footnote{
For partitions with more than one part (all except for `6'), we have tested both adjacent and distant sets of qubits for the different parts. However, this distinction did not lead to any difference in the results, which we therefore do not differentiate in Figs.~\ref{fig:entanglement_partition_assignment}.
    },
e.g., 33, we prepare W or GHZ states, as in Eq.~\eqref{eq:partitioned_GHZW_states}, on IBM machines and measure 1000 shots
in each of 500 random local bases (recall Fig.~\ref{fig:circuit}) to obtain the corresponding random correlator data (App.~\ref{appendix:UMAP}).
We feed this data to the classification pipeline, which embeds it into the same two-dimensional plane identified during training, and, consequently, associates it with a specific partition.
By comparing our data (circles and filled crosses) to the embedding of the training dataset (background clusters of points) in Fig.~\ref{fig:entanglement_partition_assignment}, we observe
correct assignment for almost all GHZ and W states prepared on the hardware.
The accuracy of the classification is rather remarkable, if one remembers that the model was trained with pure states only, whereas  states from the actual quantum devices are inevitably mixed.
To cross-check these results, we repeated the same analysis on other IBM quantum devices, namely  
\cairo, \hanoi, \auckland, \mumbai, and observed the same behaviour.

The only exception are the data for the fully entangled GHZ states---i.e., partition 6---whose correct classification crucially depends on the circuit used for their preparation. States prepared with the `ladder' circuit (see App.~\ref{appendix:GHZ_preparation}) are, up to one state, correctly assigned to the cloud, in the embedding space, that represents partition 6, while states prepared with the `star' circuit are consistently assigned to partition 33 (maroon crosses against the yellow cloud)
\footnote{
Note that by using the `ladder' circuit we could occasionally (for individual realizations but not on average over the randomly selected sets of connected qubits) also witness GHZ-type entanglement up to six qubits, which is not the case for the `star' circuit results shown in Fig.~\ref{fig:witness_scaling}.}.
The difference between these preparation circuits resides in the qubits used for the controlled gates: for the `star' circuit, all controlled gates depend on the same control qubit, which is therefore exposed to more noise (from SWAP operations and cross-talk) than the qubits in the `ladder' circuit.
Since all other states are correctly assigned to their partitions, the hardware seems not able to generate a six-partite entangled state with the `star' circuit.

\section{Entanglement Certification for Mixed States}
\label{subsec:result_entanglementcertificationmixedstates}

In Sec.~\ref{sec:entanglementwitnesses} we found that the state-specific EWs do not detect GHZ- and W-type entanglement for registers larger than four qubits.
In contrast, our entanglement classification pipeline is more robust to noise, and we are able to correctly classify entanglement partitions (with high statistical accuracy) of mixed entangled states of six qubits.
Notably, as discussed above, GHZ states prepared with the `star' circuit are classified as biseparable, while states from the `ladder' remain genuinely multipartite entangled. In line with scenario 2) discussed in Sec.~\ref{sec:entanglementwitnesses}, the former preparation circuit is apparently plagued by too much noise to generate entanglement across the entire register.
In the following, we therefore prepare states in a predefined entanglement partition, and study the latter's resilience against mixing. We do so by training a UMAP embedding on a set of random states that we generate and progressively mix, by continuously increasing the level of unbiased classical noise in the quantum state, thus interpolating between a pure and the maximally mixed state.

In practice, we consider a global \emph{depolarizing} channel \cite{Nielsen_2012}, i.e., we create states of the form \cite{Werner1989mixedstates}
\begin{equation}\label{eq:ent_class:random_mixed_part_state}
	\rho = (1-p)\ketbra{\psi}{\psi} + p \cdot \mathbb{1}/2^N ,
\end{equation}
where $\ket{\psi}$ is a random pure state as given in Eq.~\eqref{eq:ent_class:random_pure_part_state}, and $p$ is the mixing parameter.
We numerically generate one such random mixed state for each of 7000 regularly spaced values $p\in [ 0,1 ]$. 
For each of the resulting 7000 states we calculate the corresponding random correlator data (App.~\ref{appendix:UMAP}) and train a new UMAP embedding (again using a $90\%-10\%$ train-test split).

\begin{figure*}
  \includegraphics[width=\textwidth]{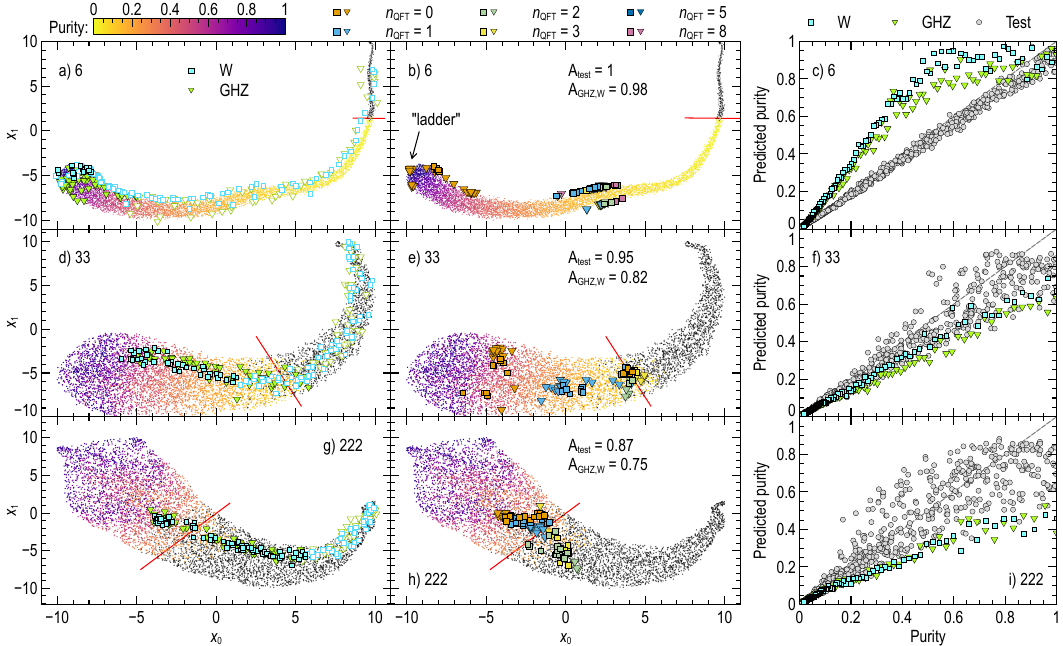}
 \caption{
 Comparison between the two-dimensional embeddings of random correlation measurement data simulated for $N=6$ qubit W and GHZ states (left column), Eq.~\eqref{eq:partitioned_GHZW_states}, and data generated on \texttt{ibmq\_ehningen} (middle column), for different partitions (from the first to the third row, the specific partition is indicated next to the panel label). 
 Each plot in the left and middle columns
 shows a serpentine-shaped cloud of points that results from training the UMAP algorithm with 7000 random states of the given partition, at  variable mixing level $p \in [0;1]$, see Eq.~\eqref{eq:ent_class:random_mixed_part_state}. 
 The $x_0$ and $x_1$ axes describe the two dimensions identified by
 the UMAP algorithm, and have no obvious physical interpretation.
 The colour of the points which constitute the serpentine-like cloud indicates the purity of the entangled states in the training dataset, 
 whereas PPT states are identified by their vanishing partition log-negativity (Sec.~\ref{subsec:result_entanglementcertificationmixedstates}) and coloured in black.
 These two categories are separated by a decision boundary (solid red line). 
 On top of the serpentines, mixed states of W and GHZ type are represented, respectively, as squares or triangles.
 In the left column, filled symbols correspond to
 entangled states, as detected by the appropriate witness (extended for the specific partition, see Sec.~\ref{subsec:simulated_WGHZ_states}). 
 For states created on \texttt{ibmq\_ehningen}, in the middle column, 
 the colour encodes the number $n_\text{QFT}$ of applications of $\mathrm{QFT}_P \cdot \mathrm{QFT}^\dagger_P$ (see Sec.~\ref{subsec:mixed-states-noisy-hardware}). 
 In the middle column we also report the accuracy scores, Eq.~\eqref{eq:accuracy} for the classification into PPT and entangled states (see Sec.~\ref{subsec:separating_entangled_states_and_ppt_states}), for states in the test dataset ($A_{\mathrm{test}}$), as well as for simulated W and GHZ-type mixed states.
 The purity of a given state can be estimated from the purity of the twenty closest states in the embedding, as we describe in more detail in Sec.~\ref{sec:purity-estimation}.
 In the right column we compare estimated and true purity for the test dataset (gray), and for the simulated W (blue) and GHZ (green) states.
	}
	\label{fig:mixednessTransition}
\end{figure*}

The resulting two-dimensional embeddings of the training datasets are shown in Fig.~\ref{fig:mixednessTransition} for three exemplary partitions (6, 33, 222),
while the remaining partitions are shown in Fig.~\ref{fig:mixedness_transition_all_partitions} in the appendix.
These embeddings take the form of serpentine-like shapes, with a colourful main body and a black tail. What sets these colours apart are the entanglement properties of the associated states.
To certify that a mixed state is entangled on a given partition $P$, we make use of the \textit{partition log-negativity} $E_\mathcal{N}^{P}$, an entanglement quantifier introduced in \cite{brunner_inprep} and defined in App.~\ref{appendix:partition_log_negatitivy}.
When $E^P_{\mathcal{N}} (\rho) > 0$, the state
$\rho$ is certainly entangled on each part $P_i$, whereas, when $E^P_\mathcal{N}(\rho) = 0$, there is at least one part $P_i$ where the state can either be entangled or separable.
By construction of the partition log-negativity, these latter states have a positive partial transpose \cite{bengtsson2017geometry} on at least one part $P_i$, and for this reason we call them PPT states (for the specific partition $P$). Because the PPT states can still be entangled, they form a proper superset of the set of separable states \cite{bengtsson2017geometry}.
These are the black points in Fig.~\ref{fig:mixednessTransition}.
We colour all other points, which are the multipartite entangled states with $E^P_\mathcal{N} > 0$, according to the purity
\begin{equation}\label{eq:purity}
\gamma = \tr \rho^2
\end{equation}
of the corresponding random mixed states [Eq.~\eqref{eq:ent_class:random_mixed_part_state}].
Since these states have a negative partial transpose on each part, we call them NPT states (again, for the specific partition $P$).

The analysis of the serpentine-like sets for the different partitions considered in Fig.~\ref{fig:mixednessTransition} suggests the following two observations:
\begin{enumerate}
    \item As we concluded in Sec.~\ref{sec:entanglementwitnesses}, if we increasingly mix a pure state, as in Eq.~\eqref{eq:ent_class:random_mixed_part_state}, we will cross the boundary into the set of separable states \cite{bengtsson2017geometry}, which belong to the PPT states.
    The distinction between PPT and the entangled NPT states is therefore present in the training dataset, and is correctly picked up by the UMAP embedding (black vs.~colourful points). Notice that the probability of finding a PPT state within the whole Hilbert space decreases fast with the number of qubits \cite{zyczkowski_volume_1998,Zyczkowski1999volume,bengtsson2017geometry}.
    In practice, we observe that the proportion of random mixed states identified by our algorithm as PPT (black dots in the left and middle panels of Fig.~\ref{fig:mixednessTransition}) decreases
    with increasing size of the largest part in the partition.
    \item The purity increases approximately monotonically and continuously along the serpentine and, therefore, the UMAP embedding is able to qualitatively capture the transition from pure to increasingly mixed quantum states.
    This implies that we can gauge, at least qualitatively, how mixed states are based on their position along the serpentine.
\end{enumerate}
In the next subsections we further elaborate on the first point, while the discussion of Sec.~\ref{sec:purity-estimation} addresses the second.

\subsection{Separating NPT and PPT states}
\label{subsec:separating_entangled_states_and_ppt_states}

While in Sec.~\ref{sec:results_entanglementpartitioncertification} we built a pipeline to associate entangled states with their partition, we now want to certify whether unseen data originates from NPT or PPT states on a given partition.
Since the embedded points from these two classes of states are well separated, as visible in Fig.~\ref{fig:mixednessTransition},
we can train, for each examined partition, a binary logistic regression classifier (see App.~\ref{appendix:UMAP}) on the embedded training dataset, with the aim of discriminating images of NPT states, with $E^P_\mathcal{N}>0$ [Eq.~\eqref{eq:ent_class:partition_log_negativity}], from those of  PPT states, with $E^P_\mathcal{N}=0$.
What we compute are linear---in the embedding plane---\emph{decision boundaries}, traced in red in the left and middle panels of Fig.~\ref{fig:mixednessTransition}, as well as in Fig.~\ref{fig:mixedness_transition_all_partitions}.
In contrast to the UMAP algorithm, which only receives the second moments of the $k$-point correlators as input data, logistic regression is a supervised learning technique, and relies on the (vanishing or finite) value of $E^P_\mathcal{N}$ associated with each point in the embedded plane.

This new entanglement certification pipeline addresses the second scenario envisioned in Sec.~\ref{sec:entanglementwitnesses}. Once the partition of the target entangled state is fixed, we can certify whether the mixed state prepared on noisy quantum hardware has remained within the partition, or has become further separable, and therefore PPT. To check the accuracy of the pipeline's prediction, we apply it to the test dataset that we generated together with the training dataset.
All partitions except 
$222$ ($A_\mathrm{test} = 0.87$) achieve high accuracy scores, Eq.~\eqref{eq:accuracy}, $A_\mathrm{test} \geq 0.94$, as reported in Fig.~\ref{fig:mixednessTransition} (middle panels) and \ref{fig:mixedness_transition_all_partitions} (right panels).

\subsection{Depolarized GHZ and W states}
\label{subsec:simulated_WGHZ_states}
Our final goal is to certify whether W or GHZ states prepared on IBM quantum hardware are entangled with respect to the expected partition. So far, we have shown that our entanglement certification pipeline is reliable on a dataset of mixed random states, which does not include W and GHZ states: since these two states are individual elements in state space, the probability of randomly drawing them for the training dataset vanishes.

We have therefore checked how the pipeline above performs on a dataset of \emph{depolarized GHZ and W states}, generated by plugging the GHZ or W states, Eq.~\eqref{eq:partitioned_GHZW_states}, into Eq.~\eqref{eq:ent_class:random_mixed_part_state}, for 200 evenly spaced values of $p \in [0,1]$.
Because we simulated these states on the computer, we know if they are PPT or NPT (by directly calculating their partition log-negativity), and with this information we can estimate the accuracy of our certification pipeline on this set of states.
Excluding the trivial case of the fully separable partition 111111,
the accuracy score $A_\text{GHZ,W}$ for the depolarized GHZ and W states is on average $12 \%$ lower than the accuracy score $A_\text{test}$ on the test dataset of mixed random states.
Considering that, as stated above, GHZ and W states do not belong to the training dataset, and that UMAP itself is not trained to detect sharp boundaries between NPT and PPT states, but rather to order states according to their degree of mixedness, we find the accuracy scores for both the test dataset and the depolarized GHZ and W states rather promising.

How do the predictions of the entanglement certification pipeline compare to the state-specific witnesses introduced in Sec.~\ref{sec:entanglementwitnesses}?
In the left panels of Figs.~\ref{fig:mixednessTransition} and Fig.~\ref{fig:mixedness_transition_all_partitions} we plot the embedded points from the depolarized GHZ (blue squares) and W (green triangles) states. Filled symbols indicate those states where all parts of the partition are witnessed as GHZ- or W-type entangled, according to the respective witness, otherwise at least one of the parts was not witnessed as entangled (empty symbols).

For partition 6, shown in Fig.~\ref{fig:mixednessTransition} a), we see that only the depolarized W states with high purity ($\gamma \gtrsim 0.8 $) are witnessed as entangled, while depolarized GHZ states with much lower purity ($\gamma \gtrsim 0.4$) are still witnessed as entangled. Indeed, state-based witnesses tolerate only so much noise---a threshold for $p$ in Eq.~\eqref{eq:ent_class:random_mixed_part_state}---beyond which they fail to detect the targeted entanglement type \cite{GUHNE20091}. 
This threshold scales as $1/N$ for W states, and is $1/2$ for GHZ states, which means that it becomes harder to witness W states on larger quantum registers. 
This translates in numerous empty symbols in Fig.~\ref{fig:mixednessTransition} a), starting from the pure end of the serpentine and up to the red boundary, associated with depolarized GHZ and W states that are \emph{not} witnessed as entangled, yet our pipeline certifies, with an accuracy score of $A_\text{GHZ,W} = 0.96$, as NPT, and hence certainly entangled.
We conclude, therefore, that our certification pipeline is significantly more robust against noise than state-specific witnesses.
As for the empty points on the black tail, they are classified as PPT states, and they are not witnessed as entangled, which implies that we cannot say anything about their entanglement properties.

If we partition the quantum register as 33, see Fig.~\ref{fig:mixednessTransition} d), we observe that the PPT states cover a larger portion of the serpentine, and that more states are witnessed as entangled, consistently with the scaling with $N$ of the noise sensitivity of the witnesses discussed above. As before, a few depolarized GHZ and W states, which are missed by the witnesses, are mapped to the NPT side of the decision boundary.
This situation changes completely if we further partition the register in 222: here the witnesses certify entanglement even though the states are mapped to the PPT side of the decision boundary. Indeed, while states with non-vanishing $E^P_\mathcal{N}$ are certainly entangled, other entangled states are PPT. Entanglement in PPT states is called \emph{bound}, because it cannot be distilled, and is in general difficult to detect with witnesses \cite{Bae_bound_2009,Huber2019,hiesmayr_free_2021}.

The different trends for partitions with larger or smaller parts 
can be observed equally well for the other partitions shown in Fig.~\ref{fig:mixedness_transition_all_partitions}.

\subsection{GHZ and W states on noisy quantum hardware}\label{subsec:mixed-states-noisy-hardware}
In Sec.~\ref{sec:entanglementwitnesses} we prepared W and GHZ states on quantum hardware from IBM, and concluded that we could not witness entanglement for registers larger than four qubits, cf.~Fig.~\ref{fig:witness_scaling}. In Sec.~\ref{sec:results_entanglementpartitioncertification}, however, we could correctly associate the (random measurements from) states prepared on IBM's hardware to the partition intended in their preparation, see Fig.~\ref{fig:entanglement_partition_assignment}.
What can we learn about the mixing of the states prepared on the quantum hardware? Can we certify that they remain entangled on the intended partition?

In the middle column of Fig.~\ref{fig:mixednessTransition} we embed the random correlator dataset of the partitioned GHZ and W states [Eq.~\eqref{eq:partitioned_GHZW_states}] prepared and measured on \texttt{ibmq\_ehningen}.
We calculate the correlators for each data point from 500 random unitary settings and 1000 shots for each of them. 
To artificially increase noise in the state preparation, without influencing the entanglement partition, we apply $n_\mathrm{QFT}$ concatenations $\mathrm{QFT}_P \cdot \mathrm{QFT}_P^\dagger$ of a partitioned quantum Fourier transform $\mathrm{QFT}_P = \mathrm{QFT}_{P_1} \otimes \dots \otimes \mathrm{QFT}_{P_r}$ followed by its inverse, on the parts $P_i$ of the considered partition $P$.
Without hardware errors and noise, this operation should leave the prepared states unchanged. 
However, since the quantum Fourier transform is implemented by a relatively complex circuit~\cite{qiskit2021}, we expect the state of the quantum register to become more mixed when prepared on the real and noisy hardware.

In Fig.~\ref{fig:mixednessTransition} (middle panel) we colour the embedded data from \texttt{ibmq\_ehningen} according to the number $n_\mathrm{QFT} = 0,1,2,3,5,8$ of applied $\mathrm{QFT}_P \cdot \mathrm{QFT}^\dagger_P$.
In general we confirm that more applications of $\mathrm{QFT}_P \cdot \mathrm{QFT}^\dagger_P$ render the quantum states more mixed, since their embeddings move closer to the maximally mixed end of the serpentine.
For partition 6, all GHZ and W states prepared on the IBM machine are certified as entangled states, since  they fall on the NPT side of the decision boundary.
Similar considerations as for the depolarized GHZ and W states hold for partitions with smaller parts, like 33 and 222. There, the portion of PPT states is much larger, and less noise is necessary to slip onto the PPT side of the decision boundary. In these cases (e.g., in partition 33 for $n_\mathrm{QFT} = 3$) it is not possible to certify entanglement anymore.
Even clearer is the case of partition 222, where for $n_\mathrm{QFT}>1$ almost all states are located in the PPT domain.

Because of finite resources on the quantum hardware,
we did not obtain data for partitions other than the three presented here.
However, from the results on the depolarized GHZ and W states, shown in Fig.~\ref{fig:mixedness_transition_all_partitions}, we do not expect to see deviations from the observations above.

\section{UMAP-based purity estimation}\label{sec:purity-estimation}

As we commented in the previous section upon inspection of the colouring of the serpentines in Fig.~\ref{fig:mixednessTransition}, the UMAP algorithm orders the embedded points of the training dataset according to their purity. Encouraged by this observation, we 
propose using the embedding of the training dataset to estimate the purity of unseen quantum states.
More specifically, we
determine those states from the training dataset associated with the $20$ points closest to the embedding of the unseen quantum state. We then average the (known) purities $\gamma$ [see. Eq.~\eqref{eq:purity}] of these $20$ states, and use this as an estimate of the unseen state's purity.

To check the validity of this method, we rely again on the test dataset, consisting of states with known purity.
In the right column of Fig.~\ref{fig:mixednessTransition} we compare, for each considered partition, estimated and true purities, while
in Fig.~\ref{fig:mixedness_transition_all_partitions} in the appendix we show the same comparison for all other partitions.
For 6, 15, 114, 1113, 11112, 111111, and 24, we obtain a narrow, linear relation between predictions and actual values. A linear regression yields correlation coefficients above $0.98$ for these partitions. Based on these linear regression models, we can obtain faithful predictions of the purity of new random states that the algorithm had not seen during training.
In turn, partitions 222, 33, 1122, and 123 show a substantial variance in the predictions, such that a reliable estimation of $\gamma$ for new data is not possible for these partitions.

In the same panels, in addition to the predictions on the test dataset, we show the prediction of the purity of the simulated depolarized W and GHZ states used in Sec.~\ref{subsec:result_entanglementcertificationmixedstates}.
While the embedding of these states is correctly ordered with increasing purity $\gamma$, significant deviations between actual and predicted values make quantitative predictions of the purity with this method unreliable.
Furthermore, for partitions with a single dominating part, i.e., 6 and 15, 
the predicted purities attain unit purity earlier than the actual purities, and from there settle on a plateau
[see Fig.~\ref{fig:mixednessTransition} c) and Fig.~\ref{fig:mixedness_transition_all_partitions} a)].
For partitions 114, 1113, 11112, 111111, and 24, instead, we identify a quite accurate linear dependence between predicted and actual $\gamma$; however, the actual values are generally underestimated. The predictions for the remaining three partitions (33, 222, 123) are subject to quite strong fluctuations.

These results suggest that, in its present form, the embedding provided by UMAP cannot be employed to reliably estimate the purity of unseen quantum states.
In particular, further investigation is necessary to understand why this estimation procedure performs worse on registers with equal parts, and why the predicted purity values of depolarized GHZ and W states reach a plateau for some partitions, e.g., Fig.~\ref{fig:mixednessTransition}~(c).
An interesting avenue are embedding spaces of dimension larger than two, which were partially explored already in \cite{brunner_thesis}. In these spaces, the algorithm may represent the progressive mixing in the family of states, defined by~\eqref{eq:ent_class:random_mixed_part_state}, more faithfully, which would also influence the estimation of the purity.

\section{Conclusions}
\label{sec:conclusions}

The question that guided our investigation is: Which entangled states can we prepare on current quantum hardware, and (how) can we certify the entanglement of the actually prepared states?

The measurement of entanglement witnesses confirmed that we did consistently prepare, on several IBM Quantum System One devices, W and GHZ states on registers of up to four qubits, but not larger (see Fig.~\ref{fig:witness_scaling}).
We observe---in agreement with earlier work~\cite{woitzik2023energy}---a striking deviation between hardware results and their simulations (see Fig.~\ref{fig:witness_noise_model}), likely due to incomplete modelling of the noise sources in the hardware.
We put forward two hypotheses to explain the failure of the witnesses: first, the register may have been prepared in an entangled state, albeit one sufficiently distinct from the targeted state; second, the noise in the preparation circuit may have erased any entanglement in the state.

To better resolve these two possibilities, we have developed two machine learning pipelines, for entanglement \emph{classification} and \emph{certification}, to, respectively, determine which partition underlies the entanglement of the prepared state, and to verify up to which noise level that state is still entangled.
Let us stress here that every statement stemming from the outcome of a machine-learning algorithm is statistical in nature, i.e., we are making a \emph{prediction} with a certain accuracy, gauged by the accuracy scores reported throughout the text. This is in stark contrast to EWs, which, although strongly susceptible to noise, are unambiguous signatures of entanglement---a specific outcome upon measurement implies that the quantum state under study certainly bears the entanglement type of the reference state.

In contrast to the state-based witnesses,
the statistical classification pipeline
is trained to detect entanglement structures in generic states of the $N$-qubit Hilbert space,
and was here shown to be able to verify that the GHZ and W states prepared on IBM machines are entangled according to the intended partition of a six-qubit register.
We observed that states prepared by the `star' circuit (which is more prone to two-qubit gate errors than the `ladder' design, see Sec.~\ref{sec:results_entanglementpartitioncertification}) are assigned to a biseparable partition, otherwise all other investigated states are consistently mapped to the targeted partition. This suggests that, indeed, the employed hardware is not able to generate a fully entangled six-qubit state with the specific `star' circuit design, possibly because the prepared states have become too mixed. This showcases the potential of our protocol of analysing the capabilities of current error-prone hardware to prepare states with specific entanglement structures. In the future, it would be interesting to apply our method to also detect whether distinct registers have become entangled due to cross-talk~\cite{Ketterer2023} between qubits.

The certification pipeline presented in this work, instead, addresses 
the stability of entanglement partitions under increasingly strong mixing.
When we train the pipeline with mixed states of a given partition, we observe a clear boundary between NPT (entangled) and PPT states (see Sec.~\ref{subsec:separating_entangled_states_and_ppt_states}), which we can statistically discriminate, with high accuracy, by training a binary classifier.
Our certification routine becomes especially valuable if we want to entangle larger registers: for increasing number of qubits, 
the relative volume of NPT with respect to that of PPT states in Hilbert space increases,
but at the same time the witness for W states becomes more sensitive to noise, therefore failing for the majority of genuinely entangled states. 
In other words, our method can effectively certify the entanglement in increasingly larger sets of mixed states, when witnesses tend to fail.

In conclusion, 
we have found an answer to our initial question: Yes, we can create entangled states on currently available quantum hardware. As long as we stick to three or four qubits, we can target, relatively faithfully,
specific entangled pure states, like GHZ or W states, and detect them with their corresponding entanglement witnesses. By lifting the constraint of state-specificity, we can instead certify (with high statistical accuracy) entanglement in mixed states of larger registers.
To further increase the accuracy of the certification pipeline, especially on W and GHZ states, it may be of interest to retrain the machine learning models on families of entangled states, e.g., on matrix product states of fixed bond dimension, which automatically include W and GHZ states \cite{schollwock2011density, cirac2021matrix}.
The use of tensor-network methods would, in addition, allow us to investigate the performance of our models when trained on entangled states with low bond dimension, which can be efficiently simulated on classical computers.

\section*{Acknowledgements}

The authors thanks David Amaro and Luuk Coopmans for feedback on the manuscript.
The authors acknowledge support by the state of Baden-Württemberg through bwHPC (High Performance Computing, Baden-Württemberg), and funding by the Deutsche Forschungsgemeinschaft (DFG, German Research Foundation)|Grant No. INST 40/575-1 FUGG (JUSTUS 2 cluster).
	A.~J.~C.~F.\ acknowledges the Konrad-Adenauer-Foundation for financial support.
	E.~G.~C.\ acknowledges support from the Georg H.~Endress foundation and from the project ``SiQuRe-II'' (Kompetenzzentrum Quantencomputing Baden-W\"urttemberg).
	Funded by the Ministerium für Wirtschaft, Arbeit und Tourismus of the State of Baden-Württemberg.
J.~S., H.~K., S.~S., and J.~B.\
acknowledge the support from the National Research Foundation of Korea (NRF-2021R1A2C2006309, NRF-2020K2A9A2A15000061, RS-2024-00408613, RS-2023-00257994) and the Institute for Information \& Communication Technology Promotion (IITP) (RS-2023-00229524).

\appendix

\section{Witnesses}
\label{appendix:witnesses}

To determine the expectation value of a witness $\langle \mathcal{W} \rangle$ on the given hardware, we need to expand $\mathcal{W}$ as a sum of strings of local operators $O_1^j \otimes ...\otimes O_N^j$, whose expectation values can be measured:
\begin{equation}
    \mathcal{W} = \sum_j c_j O_1^j \otimes ...\otimes O_N^j,
\end{equation}
with $c_j$ the necessary expansion coefficients to reconstruct $\langle \mathcal{W} \rangle$ from the measurement outcomes.
Previous work~\cite{guhne2007toolbox} has shown that the minimal number of local operators required to determine $\mathcal{W}_{\mathrm{W}_N}$ and $\mathcal{W}_{\mathrm{GHZ}_N}$, is $2N-1$ and $N+1$, respectively, and therefore scales linearly with the number of qubits. 

For GHZ states, these local measurement operators are $(Z)^{\otimes N}$ and equidistant rotations in the $x$-$y$ plane of the Bloch sphere:
\begin{equation}\label{eq:M_k_GHZ}
    \mathcal{M}_k = \left[ \cos(\frac{k\pi}{N})X + \sin(\frac{k\pi}{N})Y \right] ^{\otimes N} , k = 1,...,N .
\end{equation}
The expectation value of $\mathcal{W}_{\mathrm{GHZ}_N}$ can then be reconstructed by inserting in Eq.~\eqref{eq:ghz_witness} the following equality \cite{guhne2007toolbox}:
\begin{align}
    2 \dyad{\mathrm{GHZ}_N} = & \frac{1}{2^N} \left[ (\mathbb{I}+Z)^{\otimes N} +  (\mathbb{I}-Z)^{\otimes N} \right] \nonumber \\
    & + \frac{1}{N} \sum_{k=1}^N (-1)^k \mathcal{M}_k.
\end{align}

For W states, instead, we measure the operators
\begin{equation}\label{eq:M_k_W}
    \mathcal{M}_k^{\pm}(\mathcal{O}) = \left[ \cos(\alpha_k) \mathcal{O} \pm \sin(\alpha_k)(\mathbb{1} + Z) \right] ^{\otimes N} , 
\end{equation}
for both choices $\mathcal{O}= X$ or $Y$, and for the measurement directions set by $\alpha_k = (\pi/2) (k+1)/(L+1)$, $k = 1,...,L$, where $L = \lfloor (N-1)/2\rfloor$. Since
$\mathcal{M}_L^+ = \mathcal{M}_L^-$,
we need in total $2N-1$ settings.
As for the GHZ state, the reconstruction of $\mathcal{W}_{\mathrm{W}_N}$ requires inserting in Eq.~\eqref{eq:w_witness} the following equality \cite{guhne2007toolbox}:
\begin{align}\label{app:eq:W-witness-decomposition}
	N 2^N \dyad{\text{W}_N} = & \sum_{k=1}^N (\mathbb{I}+Z)^{\otimes k-1}(\mathbb{I}-Z)(\mathbb{I}+Z)^{\otimes N-k}  \nonumber \\
	& - 4 \gamma_0 \left( \mathbb{I} + Z\right)^{\otimes N} - 2 \gamma_N \left( X^{\otimes N} + Y^{\otimes N}\right) \nonumber \\
	& + 2\sum_{j=1}^L \chi_j [\mathcal{Q}_j(X)+\mathcal{Q}_j(Y)] ,
\end{align}
where, for compactness, we have introduced the operators $\mathcal{Q}_k(\mathcal{O}) = [\mathcal{M}_k^{+}(\mathcal{O}) + (-1)^N \mathcal{M}_k^{-}(\mathcal{O})]/2$. The weights $\chi_j$ depend on the set of angles $\alpha_k$ in Eq.~\eqref{eq:M_k_W}, and are found by solving the system of equations
\begin{equation}
	\sum_{j=1}^L \chi_j (\cos \alpha_j)^k (\sin\alpha_j)^{N-k} = \delta_{k,2} \, ,
\end{equation}
for $k$ even and $2 \leq k < N$. From the weights we can calculate the remaining coefficients, i.e., $\gamma_0 = \sum_{j=1}^L \chi_j (\sin \alpha_j)^N $, and $\gamma_N = \sum_{j=1}^L \chi_j (\cos \alpha_j)^N $ if $N$ is odd, otherwise $\gamma_N = 0$ [implying that no measurement of $ X^{\otimes N}$ or $ Y^{\otimes N}$ is needed, see Eq.~\eqref{app:eq:W-witness-decomposition}].

\section{Circuits for the preparation of GHZ states}
\label{appendix:GHZ_preparation}
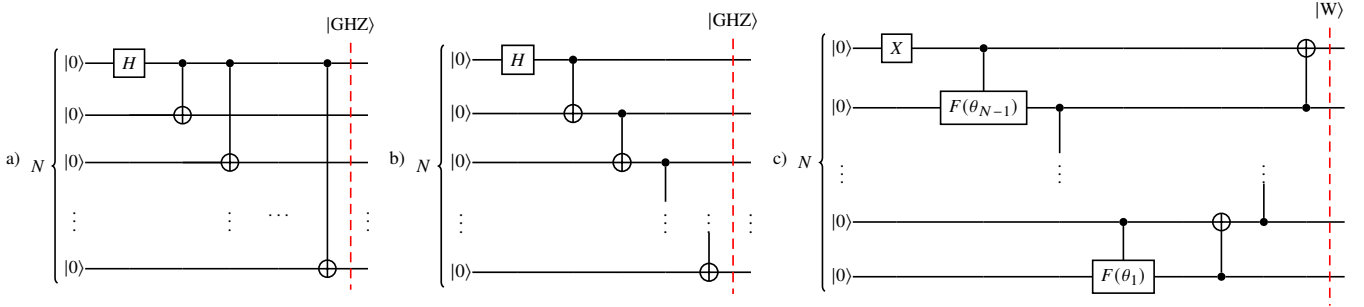
\begin{figure*}
\begin{center}
\begin{adjustbox}{width=.28\textwidth}
a)
\begin{quantikz} \lstick[wires=5]{$N$}
	\ket{0}  & \gate{H} & \ctrl{1} & \ctrl{2} & \qw  & \ctrl{4} \slice{$\ket{\text{GHZ}}$} & \qw \\
	\ket{0}  & \qw & \targ & \qw &  \qw & \qw & \qw &  \qw\\
	\ket{0}  & \qw & \qw & \targ &  \qw & \qw & \qw & \qw \\
	\vdots 	& 		& 	 &  \vdots & \dots & & \vdots\\
	\ket{0}  & \qw & \qw & \qw &  \qw & \targ{} &  \qw 
\end{quantikz}
\end{adjustbox}
\begin{adjustbox}{width=.28\textwidth}
b)
\begin{quantikz} \lstick[wires=5]{$N$}
	\ket{0}  & \gate{H} & \ctrl{1} & \qw & \qw  & \qw \slice{$\ket{\text{GHZ}}$} &   \qw  \\
	\ket{0}  & \qw & \targ{} & \ctrl{1} & \qw & \qw     &  \qw \\
	\ket{0}  & \qw & \qw & \targ{} & \ctrl{1} & \qw     &  \qw \\
	\vdots 	 & 	   &	 & 		   &  \vdots  & \vdots & \vdots  \\
	\ket{0}  & \qw & \qw & \qw     &  \qw    & \targ{} \vqw{-1}  &  \qw
\end{quantikz}
\end{adjustbox}
\begin{adjustbox}{width=.43\textwidth}
c)
\begin{quantikz} \lstick[wires=5]{$N$}
	\ket{0}  & \gate{X} & \ctrl{1} & \qw &    \qw &  \qw &  \qw &  \qw &  \targ{} \slice{$\ket{\mathrm{W}}$} &  \qw  \\
	\ket{0}  & \qw & \gate{F(\theta_{N-1})} & \ctrl{1} & \qw & \qw &  \qw &  \qw &  \ctrl{-1} &   \qw \\
	\vdots 	& 		& 		  &  \vdots & &  & & \vdots & \\
	\ket{0}  & \qw & \qw & \qw &  \ctrl{1} &   \qw & \targ{} & \ctrl{-1} &  \qw &   \qw \\
	\ket{0}  & \qw & \qw & \qw & \gate{F(\theta_1)} & \qw & \ctrl{-1} &  \qw  &  \qw & \qw
\end{quantikz}
\end{adjustbox}
\caption{
Sketches of the circuits for the preparation, on gate-based hardware like IBM's, of the GHZ state using a) the `star' configuration or b) the `ladder' configuration, and of c) the W state.
On the heavy hexagonal topology \cite{zhang2022highperformance} of IBM machines, the `star' circuit requires additional SWAP gates for each applied CNOT gate
(see Sec.~\ref{sec:results_entanglementpartitioncertification}), and is therefore more susceptible to hardware noise than the `ladder' circuit. The preparation of the W state makes use of the controlled rotations defined \cite{diker2016deterministic} as
$F(\theta) = R_y(\theta) Z R_y(-\theta)$, with angles $\theta_k = \arccos( \sqrt{k}/\sqrt{k+1} ) $ for $k=1,\ldots,N-1$.
}
\label{app:fig:ghz_circuits}
\end{center}
\end{figure*}

In Fig.~\ref{app:fig:ghz_circuits} we show the two possible circuits (`star' \cite{Gorbachev2000} and `ladder') that can be used, in principle equivalently, to prepare a GHZ state.

\section{Partition log-negativity}
\label{appendix:partition_log_negatitivy}

The logarithmic negativity, introduced in \cite{vidal_computable_2002}, is a bipartite entanglement monotone for mixed states \cite{plenio_logarithmic_2005} that is based on the positive-partial-transpose (PPT) criterion \cite{peres_separability_1996,horodecki_separability_1996}. For a given bipartition $[A,\bar{A}]$, with $\bar{A}$ the complement to $A$, the logarithmic negativity reads
\begin{equation}
	E_{\mathcal{N}, A} (\rho) = \log_2  \Vert \rho^{\Gamma_A} \Vert_1 \,,
\end{equation}
where $\Vert X \Vert_1 = \tr \sqrt{ X^\dagger X}$ is the trace norm, and $\rho^{\Gamma_A}$ describes the partial transpose of $\rho$ with respect to $A$.
Based on the logarithmic negativity, the \textit{partition log-negativity} for a given $P \in \mathbb{P}$ is defined \cite{brunner_inprep} as
\begin{equation}\label{eq:ent_class:partition_log_negativity}
E^P_\mathcal{N}(\rho) = \prod_{P_i \in P  : |P_i| > 1} \tilde{E}_\mathcal{N} (\rho_{P_i})^{|P_i|/\Vert P\Vert }\,,
\end{equation}
where $\rho_{P_i}$ is the reduced state of $\rho$ on the part $P_i$.
With $\Vert P\Vert$ we denote the number of entangled qubits in partition $P$ (i.e., not counting the qubits that separate from the others).
For example, we have $\big\Vert [[0],[1,2,3,4,5]] \big\Vert = 5$ and $\big\Vert [[0,1],[2,3,4,5]] \big\Vert = 6$.
The function $\tilde{E}_\mathcal{N}$ is the geometric mean of logarithmic negativities over all bipartitions of $\rho_{P_i}$, 
\begin{equation}
\tilde E_\mathcal{N} (\rho_{P_i}) = \left[ \prod_{\text{bipart.~$[A,\bar{A}]$ of $\rho_{P_i}$}} E_{\mathcal{N}, A} (\rho_{P_i}) \right]^{1/C} \,,
\end{equation}
where $C$ is the number of such bipartitions.
Hence, the partition log-negativity in Eq.~\eqref{eq:ent_class:partition_log_negativity} is nothing but the (weighted) geometric mean over the geometric means $\tilde E_\mathcal{N} (\rho_{P_i})$ of all bipartite logarithmic negativities of the parts of $P$.
Due to the geometric mean, $E^P_{\mathcal{N}} (\rho) > 0$ implies that $\rho$ has non-vanishing average logarithmic negativity on each and any part $P_i$ of $P$.
Since the state of any part $P_i$ of $P$ can be entangled, and yet have zero logarithmic negativity, there exist entangled states with $E^P_\mathcal{N}(\rho) = 0$.

\section{Dataset and non-linear dimensionality reduction}
\label{appendix:UMAP}

\subsection{The dataset}
\label{app:UMAP_thedataset}

To characterize the multipartite entanglement structure of the state $\ket{\psi}$ generated by the quantum circuit in Fig.~\ref{fig:circuit}, we consider the distributions of correlators in random single-qubit bases. 
These distributions have already been proposed for entanglement detection in~\cite{van_enk_measuring_2012,elben_statistical_2019,knips_multipartite_2020,ketterer_statistically_2022,wyderka2023complete}.
To this end, we measure $k$-point correlators in the $z$ direction (i.e., the computational basis) after independent random unitary rotations $R_1,\dots, R_N$ (see pre-measurement rotations in Fig.~\ref{fig:circuit}) of the individual qubits of the prepared state $\ket{\psi}$:
\begin{equation}\label{eq:ent_class:random_correlator}
	c_{m_1,\dots, m_k} = \expv{R_{m_1}^\dagger Z_{m_1} R_{m_1} \dots R_{m_k}^\dagger Z_{m_k} R_{m_k}} \,,
\end{equation}
where $Z_m$ is the Pauli-$Z$ operator on qubit $m$.
Note that in Eq.~\eqref{eq:ent_class:random_correlator} we consider the standard instead of the connected correlators employed in \cite{brunner_inprep}.
Both sets of correlators contain the same information, and, from our analyses, using either yields no appreciable difference in the results.
The distribution of $c_{m_1,\dots, m_k}$, with respect to the random sampling of Haar unitaries $R_i$, is characterized by their second moment
\begin{equation}\label{eq:ent_class:statistical_moments}
	M^{2}_{m_1, \dots, m_k} = \int dR_{m_1} \dots dR_{m_k} \left( c_{m_1, \dots, m_k} \right)^2 \,. 
\end{equation}
A similar approach was followed in \cite{brunner_many-body_2022} in a non-interacting (e.g., photonic) setting, where the first moments of the output correlations were shown to give access to coherence quantifiers of the initial state.
The dimension $D$ of the obtained dataset, i.e., the second moments of the $k$-point correlators up to order $k \leq N$, is given by $D = \sum_{k = 1}^N \binom{N}{k} = \big( 2^N -1 \big)$.
In our calculations, we estimate these second moments from $N_\mathrm{unit} = 500$ $k$-tuples of independent single-qubit Haar unitaries $(R_1, \dots, R_k)$.

\subsection{Non-linear dimensionality reduction (manifold learning)}

To extract the relevant information about the entanglement and the mixing of the quantum states, we apply the non-linear dimensionality reduction (also known as manifold learning) algorithm UMAP (\textit{uniform manifold approximation and projection}) introduced in \cite{mcinnes2018umap}.
UMAP is a widely used method for feature extraction and visualization, which has already found application in diverse fields such as biology~\cite{cao_single-cell_2019}, machine learning~\cite{diaz-papkovich_umap_2019,carter_activation_2019}  and time-series analysis~\cite{ali_timecluster_2019}.
The basic idea behind non-linear dimensionality reduction is to embed the manifold structure of a high-dimensional dataset $X = \mathbb{R}^{D}$ into a low-dimensional representation space $Y$.
The embedding $f : X \rightarrow Y$ is obtained by, first, generating a graph $G_X$ of the input data $x_i$.
The edge weights $v_{ij}$ between points $x_i, x_j$ are calculated from a suitable distance measure in $X$.
In the same way, an embedding graph $G_Y$ is defined in the low-dimensional space $Y$ with edge weights $w_{ij}$ computed from, at first, randomly initialized embedded points $y_i, y_j \in Y$.
The points $y_i$ are then optimized by minimizing a distance measure -- such as the Kullback-Leibler divergence~\cite{kullback_information_1951}, or slight variations thereof -- between the distributions $\lbrace v_{ij} \rbrace$ and $\lbrace w_{ij} \rbrace$.
The edge values $w_{ij}$ (between embedded points $y_i$ and $y_j$) of this graph are found by minimizing a distance measure -- such as the Kullback-Leibler divergence~\cite{kullback_information_1951}, or slight variations thereof -- between the distributions $\lbrace v_{ij} \rbrace$ and $\lbrace w_{ij} \rbrace$. The UMAP algorithm specifically uses
\begin{equation}
C(v,w) = \sum_{i \neq j} v_{ij} \log \frac{v_{ij}}{w_{ij}} + (1 - v_{ij}) \log \frac{1- v_{ij}}{1 - w_{ij}} \,.
\end{equation}
The precise definitions of the edge weights and metrics used by UMAP are not important for our analysis here, and can be found in \cite{mcinnes2018umap}.

In this work, we always use a two-dimensional embedding space $Y = \mathbb{R}^2$, which is suitable for visualizing the embedded data.
Like many statistical algorithms, the application of UMAP requires standardised data, i.e., data with unit variance and zero mean. 
Hence, we first calculate variance and mean of each second moment in Eq.~\eqref{eq:ent_class:statistical_moments} across the set of randomly generated states, and apply the appropriate scaling before feeding the data to the UMAP algorithm. 
Apart from the embedding dimension, UMAP has two main hyperparameters: the number $n_\mathrm{neighbors}$ of nearest neighbors of $x_i$, from which the edge weights $v_{ij}$ are determined, and a minimal distance $d_\mathrm{min}$ between any two points in the low-dimensional embedding. They are described in more detail below in App.~\ref{app:umap_hyperparameter}.

As for the scalability of the proposed scheme, the number of random unitaries to estimate, up to accuracy $\delta$, the statistical moments of the $k$-point correlators scales exponentially in $k$, as discussed in \cite{ketterer_statistically_2022,brunner_inprep}. Hence, we cannot apply the scheme to study all entanglement partitions of a large $N$-qubit systems.
However, as proposed in \cite{brunner_inprep}, one can train the model on data from a $N_0$-qubit system (with constant $N_0 < N$) and apply it to all subsets of at most $N_0$ qubits from a large $N$-qubit system. The number of these subsets scales polynomially, $\binom{N}{N_0} \sim N^{N_0}$, in $N$. In this way, we obtain a scalable protocol to analyse multipartite entanglement in all subsets of up to $N_0$ out of $N$ qubits.

As proposed in \cite{brunner_inprep}, we employ the UMAP embedding algorithm for two tasks: (1) the classification of entanglement partitions of mixed quantum states, and (2) 
the certification of the entanglement inscribed into a mixed state prepared according to a given partition.
For the first task, we use UMAP to embed the second moments of the random correlators, stemming from the various entanglement partitions, in well-separated clusters that identify those partitions (see Fig.~\ref{fig:entanglement_partition_assignment}). Based on these clusters, we train a standard decision-tree classifier (we use the implementation of scikit-learn~\cite{scikit-learn}) to classify the entanglement partitions of simulated GHZ and W states [Eq.~\eqref{eq:partitioned_GHZW_states}], and of states prepared on actual quantum hardware.

For the second task we embed data from random states of a given partition, subjected to the global depolarizing channel in Eq.~\eqref{eq:ent_class:random_mixed_part_state}, with the UMAP algorithm. The resulting low-dimensional representation
shows a clear separation between NPT and PPT states (see Sec.~\ref{subsec:separating_entangled_states_and_ppt_states} and Fig.~\ref{fig:mixednessTransition}). Based on this embedding, we train a logistic regression model as binary classifier [also implemented in scikit-learn (v1.1.2) \cite{scikit-learn}] to discriminate between NPT and PPT mixed quantum states.

\subsection{UMAP hyperparameters}
\label{app:umap_hyperparameter}

The two hyperparameters $n_\mathrm{neighbors}$ and $d_\mathrm{min}$ can be tuned to shift the focus between
local and global ``structures'' in the low-dimensional embedding. For example, a small $n_\mathrm{neighbors}$ and a small $d_\mathrm{min}$ tend to produce embeddings with many small clusters, i.e., the algorithm picks up on fine differences between the data points, and clusters together only points with a high degree of similarity, as encoded in the edge weights $v_{ij}$. In turn, large $n_\mathrm{neighbors}$ and $d_\mathrm{min}$ favours the generation of less clustered and more connected embeddings.
Note, however, that in general the choice of these hyperparameters must also take into account the size of the dataset.
For a more detailed description, see \cite{mcinnes2018umap}.

\paragraph{Hyperparameter: Classification of entanglement partitions}

To classify entanglement partitions, see Fig.~\ref{fig:entanglement_partition_assignment}, we choose the hyperparameters $n_\mathrm{neighbors} = 50$ and $d_\mathrm{min} = 0.3$.
These relatively small parameters are a suitable choice in this scenario, where we wish to discriminate a large number of different classes, the entanglement partitions, by examination of the structure of the random correlator dataset [cf. App.~\ref{app:UMAP_thedataset}].

\paragraph{Hyperparameter: Entanglement certification for mixed states}

To map out the continuous transition between a given pure state's entanglement partition and the maximally mixed state, we need to adjust the UMAP algorithm to focus more on global aspects of the dataset. More concretely, we want to prevent UMAP from splitting up the embedded dataset in different clusters. For this, we need to choose a larger number of nearest neighbours and a larger minimal distance. For the simulations of Figs.~\ref{fig:mixednessTransition} and \ref{fig:mixedness_transition_all_partitions} we used $n_\mathrm{neighbors} = 500$ and $d_\mathrm{min} = 0.6$.

\section{Certification of mixed states of the remaining partitions}
\label{app:all-snakes}

\begin{figure*}
    \includegraphics[width=\textwidth]{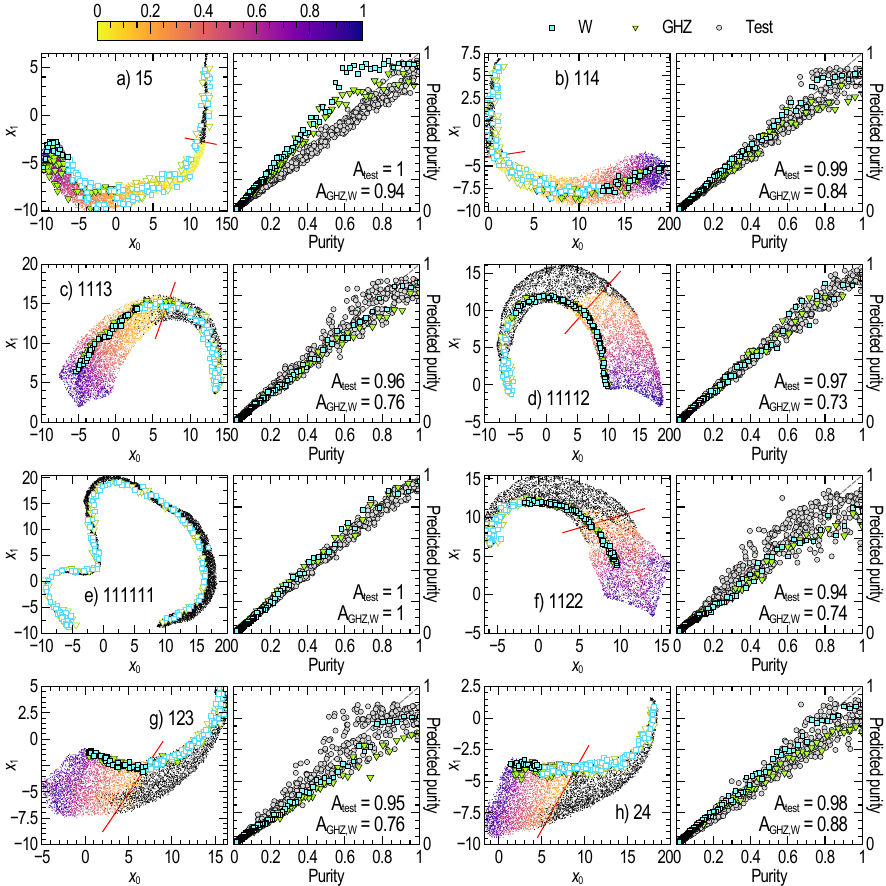}
 \caption{
 (a-h) Embedding of simulated mixed states, and purity determination accuracy, for different partitions.
 Each panel on the left shows a background serpentine that results from training the UMAP algorithm with 7000 random states, Eq.~\eqref{eq:ent_class:random_mixed_part_state}, of the indicated partition, and different mixing levels $p\in[0;1]$.
 The axes describe the two dimensions identified by the UMAP algorithm (see App.~\ref{appendix:UMAP}), and have no obvious physical interpretation.
 All graphical elements (colours and shapes) follow the same coding as in Fig.~\ref{fig:mixednessTransition}.
 In the panels on the right we report the accuracy scores, Eq.~\eqref{eq:accuracy}, for the certification of test and simulated W and GHZ states (see Sec.~\ref{subsec:separating_entangled_states_and_ppt_states}), as well as the comparison of estimated and true purity for the test dataset (gray), and for the simulated W (blue) and GHZ (green) states, see Sec.~\ref{sec:purity-estimation}.
}
\label{fig:mixedness_transition_all_partitions}
\end{figure*}

In Fig.~\ref{fig:mixednessTransition} we showed three example partitions, abbreviated 6, 33, and 222, each of which consists of equally sized parts.
In Fig.~\ref{fig:mixedness_transition_all_partitions} we present the embedding of all other partitions, i.e., partitions with parts of unequal sizes, as well as the embedding for the fully separable states.
Qualitative observations are comparable for partitions with a similar structure. 
For instance, for the partition 15 we observe the same plateau of the purity of the W states as for partition 6 [see Fig.~\ref{fig:mixednessTransition} a)].
For a more detailed comparison of the different cases, see Sec.~\ref{subsec:result_entanglementcertificationmixedstates}.

\begin{thebibliography}{87}%
\makeatletter
\providecommand \@ifxundefined [1]{%
 \@ifx{#1\undefined}
}%
\providecommand \@ifnum [1]{%
 \ifnum #1\expandafter \@firstoftwo
 \else \expandafter \@secondoftwo
 \fi
}%
\providecommand \@ifx [1]{%
 \ifx #1\expandafter \@firstoftwo
 \else \expandafter \@secondoftwo
 \fi
}%
\providecommand \natexlab [1]{#1}%
\providecommand \enquote  [1]{``#1''}%
\providecommand \bibnamefont  [1]{#1}%
\providecommand \bibfnamefont [1]{#1}%
\providecommand \citenamefont [1]{#1}%
\providecommand \href@noop [0]{\@secondoftwo}%
\providecommand \href [0]{\begingroup \@sanitize@url \@href}%
\providecommand \@href[1]{\@@startlink{#1}\@@href}%
\providecommand \@@href[1]{\endgroup#1\@@endlink}%
\providecommand \@sanitize@url [0]{\catcode `\\12\catcode `\$12\catcode
  `\&12\catcode `\#12\catcode `\^12\catcode `\_12\catcode `\%12\relax}%
\providecommand \@@startlink[1]{}%
\providecommand \@@endlink[0]{}%
\providecommand \url  [0]{\begingroup\@sanitize@url \@url }%
\providecommand \@url [1]{\endgroup\@href {#1}{\urlprefix }}%
\providecommand \urlprefix  [0]{URL }%
\providecommand \Eprint [0]{\href }%
\providecommand \doibase [0]{https://doi.org/}%
\providecommand \selectlanguage [0]{\@gobble}%
\providecommand \bibinfo  [0]{\@secondoftwo}%
\providecommand \bibfield  [0]{\@secondoftwo}%
\providecommand \translation [1]{[#1]}%
\providecommand \BibitemOpen [0]{}%
\providecommand \bibitemStop [0]{}%
\providecommand \bibitemNoStop [0]{.\EOS\space}%
\providecommand \EOS [0]{\spacefactor3000\relax}%
\providecommand \BibitemShut  [1]{\csname bibitem#1\endcsname}%
\let\auto@bib@innerbib\@empty
\bibitem [{\citenamefont {Van~den Nest}\ \emph {et~al.}(2006)\citenamefont
  {Van~den Nest}, \citenamefont {Miyake}, \citenamefont {D{\"u}r},\ and\
  \citenamefont {Briegel}}]{van2006universal}%
  \BibitemOpen
  \bibfield  {author} {\bibinfo {author} {\bibfnamefont {M.}~\bibnamefont
  {Van~den Nest}}, \bibinfo {author} {\bibfnamefont {A.}~\bibnamefont
  {Miyake}}, \bibinfo {author} {\bibfnamefont {W.}~\bibnamefont {D{\"u}r}},\
  and\ \bibinfo {author} {\bibfnamefont {H.~J.}\ \bibnamefont {Briegel}},\
  }\bibfield  {title} {\bibinfo {title} {Universal resources for
  measurement-based quantum computation},\ }\href
  {https://doi.org/10.1103/PhysRevLett.97.150504} {\bibfield  {journal}
  {\bibinfo  {journal} {Physical Review Letters}\ }\textbf {\bibinfo {volume}
  {97}},\ \bibinfo {pages} {150504} (\bibinfo {year} {2006})}\BibitemShut
  {NoStop}%
\bibitem [{\citenamefont {Markov}\ and\ \citenamefont
  {Shi}(2008)}]{markov2008simulating}%
  \BibitemOpen
  \bibfield  {author} {\bibinfo {author} {\bibfnamefont {I.~L.}\ \bibnamefont
  {Markov}}\ and\ \bibinfo {author} {\bibfnamefont {Y.}~\bibnamefont {Shi}},\
  }\bibfield  {title} {\bibinfo {title} {Simulating quantum computation by
  contracting tensor networks},\ }\href {https://doi.org/10.1137/050644756}
  {\bibfield  {journal} {\bibinfo  {journal} {SIAM Journal on Computing}\
  }\textbf {\bibinfo {volume} {38}},\ \bibinfo {pages} {963} (\bibinfo {year}
  {2008})}\BibitemShut {NoStop}%
\bibitem [{\citenamefont {Nielsen}\ and\ \citenamefont
  {Chuang}(2012)}]{Nielsen_2012}%
  \BibitemOpen
  \bibfield  {author} {\bibinfo {author} {\bibfnamefont {M.~A.}\ \bibnamefont
  {Nielsen}}\ and\ \bibinfo {author} {\bibfnamefont {I.~L.}\ \bibnamefont
  {Chuang}},\ }\href {https://doi.org/10.1017/cbo9780511976667} {\emph
  {\bibinfo {title} {Quantum Computation and Quantum Information}}}\ (\bibinfo
  {publisher} {Cambridge University Press},\ \bibinfo {year}
  {2012})\BibitemShut {NoStop}%
\bibitem [{\citenamefont {Zhou}\ \emph {et~al.}(2020)\citenamefont {Zhou},
  \citenamefont {Stoudenmire},\ and\ \citenamefont {Waintal}}]{zhou2020limits}%
  \BibitemOpen
  \bibfield  {author} {\bibinfo {author} {\bibfnamefont {Y.}~\bibnamefont
  {Zhou}}, \bibinfo {author} {\bibfnamefont {E.~M.}\ \bibnamefont
  {Stoudenmire}},\ and\ \bibinfo {author} {\bibfnamefont {X.}~\bibnamefont
  {Waintal}},\ }\bibfield  {title} {\bibinfo {title} {What {{Limits}} the
  {{Simulation}} of {{Quantum Computers}}?},\ }\href
  {https://doi.org/10.1103/PhysRevX.10.041038} {\bibfield  {journal} {\bibinfo
  {journal} {Physical Review X}\ }\textbf {\bibinfo {volume} {10}},\ \bibinfo
  {pages} {041038} (\bibinfo {year} {2020})}\BibitemShut {NoStop}%
\bibitem [{\citenamefont {Preskill}(2018)}]{preskill2018quantum}%
  \BibitemOpen
  \bibfield  {author} {\bibinfo {author} {\bibfnamefont {J.}~\bibnamefont
  {Preskill}},\ }\bibfield  {title} {\bibinfo {title} {Quantum computing in the
  nisq era and beyond},\ }\href {https://doi.org/10.22331/q-2018-08-06-79}
  {\bibfield  {journal} {\bibinfo  {journal} {Quantum}\ }\textbf {\bibinfo
  {volume} {2}},\ \bibinfo {pages} {79} (\bibinfo {year} {2018})}\BibitemShut
  {NoStop}%
\bibitem [{\citenamefont {He}\ \emph {et~al.}(2020)\citenamefont {He},
  \citenamefont {Nachman}, \citenamefont {De~Jong},\ and\ \citenamefont
  {Bauer}}]{he2020zeronoise}%
  \BibitemOpen
  \bibfield  {author} {\bibinfo {author} {\bibfnamefont {A.}~\bibnamefont
  {He}}, \bibinfo {author} {\bibfnamefont {B.}~\bibnamefont {Nachman}},
  \bibinfo {author} {\bibfnamefont {W.~A.}\ \bibnamefont {De~Jong}},\ and\
  \bibinfo {author} {\bibfnamefont {C.~W.}\ \bibnamefont {Bauer}},\ }\bibfield
  {title} {\bibinfo {title} {Zero-noise extrapolation for quantum-gate error
  mitigation with identity insertions},\ }\href
  {https://doi.org/10.1103/PhysRevA.102.012426} {\bibfield  {journal} {\bibinfo
   {journal} {Physical Review A}\ }\textbf {\bibinfo {volume} {102}},\ \bibinfo
  {pages} {012426} (\bibinfo {year} {2020})}\BibitemShut {NoStop}%
\bibitem [{\citenamefont {Magesan}\ \emph {et~al.}(2012)\citenamefont
  {Magesan}, \citenamefont {Gambetta}, \citenamefont {Johnson}, \citenamefont
  {Ryan}, \citenamefont {Chow}, \citenamefont {Merkel}, \citenamefont
  {Da~Silva}, \citenamefont {Keefe}, \citenamefont {Rothwell}, \citenamefont
  {Ohki}, \citenamefont {Ketchen},\ and\ \citenamefont
  {Steffen}}]{magesan2012efficient}%
  \BibitemOpen
  \bibfield  {author} {\bibinfo {author} {\bibfnamefont {E.}~\bibnamefont
  {Magesan}}, \bibinfo {author} {\bibfnamefont {J.~M.}\ \bibnamefont
  {Gambetta}}, \bibinfo {author} {\bibfnamefont {B.~R.}\ \bibnamefont
  {Johnson}}, \bibinfo {author} {\bibfnamefont {C.~A.}\ \bibnamefont {Ryan}},
  \bibinfo {author} {\bibfnamefont {J.~M.}\ \bibnamefont {Chow}}, \bibinfo
  {author} {\bibfnamefont {S.~T.}\ \bibnamefont {Merkel}}, \bibinfo {author}
  {\bibfnamefont {M.~P.}\ \bibnamefont {Da~Silva}}, \bibinfo {author}
  {\bibfnamefont {G.~A.}\ \bibnamefont {Keefe}}, \bibinfo {author}
  {\bibfnamefont {M.~B.}\ \bibnamefont {Rothwell}}, \bibinfo {author}
  {\bibfnamefont {T.~A.}\ \bibnamefont {Ohki}}, \bibinfo {author}
  {\bibfnamefont {M.~B.}\ \bibnamefont {Ketchen}},\ and\ \bibinfo {author}
  {\bibfnamefont {M.}~\bibnamefont {Steffen}},\ }\bibfield  {title} {\bibinfo
  {title} {Efficient {{Measurement}} of {{Quantum Gate Error}} by {{Interleaved
  Randomized Benchmarking}}},\ }\href
  {https://doi.org/10.1103/PhysRevLett.109.080505} {\bibfield  {journal}
  {\bibinfo  {journal} {Physical Review Letters}\ }\textbf {\bibinfo {volume}
  {109}},\ \bibinfo {pages} {080505} (\bibinfo {year} {2012})}\BibitemShut
  {NoStop}%
\bibitem [{\citenamefont {Nachman}\ \emph {et~al.}(2020)\citenamefont
  {Nachman}, \citenamefont {Urbanek}, \citenamefont {De~Jong},\ and\
  \citenamefont {Bauer}}]{nachman2020unfolding}%
  \BibitemOpen
  \bibfield  {author} {\bibinfo {author} {\bibfnamefont {B.}~\bibnamefont
  {Nachman}}, \bibinfo {author} {\bibfnamefont {M.}~\bibnamefont {Urbanek}},
  \bibinfo {author} {\bibfnamefont {W.~A.}\ \bibnamefont {De~Jong}},\ and\
  \bibinfo {author} {\bibfnamefont {C.~W.}\ \bibnamefont {Bauer}},\ }\bibfield
  {title} {\bibinfo {title} {Unfolding quantum computer readout noise},\ }\href
  {https://doi.org/10.1038/s41534-020-00309-7} {\bibfield  {journal} {\bibinfo
  {journal} {npj Quantum Information}\ }\textbf {\bibinfo {volume} {6}},\
  \bibinfo {pages} {84} (\bibinfo {year} {2020})}\BibitemShut {NoStop}%
\bibitem [{\citenamefont {Woitzik}\ \emph {et~al.}(2024)\citenamefont
  {Woitzik}, \citenamefont {Hoffmann}, \citenamefont {Buchleitner},\ and\
  \citenamefont {Carnio}}]{woitzik2023energy}%
  \BibitemOpen
  \bibfield  {author} {\bibinfo {author} {\bibfnamefont {A.~J.~C.}\
  \bibnamefont {Woitzik}}, \bibinfo {author} {\bibfnamefont {L.}~\bibnamefont
  {Hoffmann}}, \bibinfo {author} {\bibfnamefont {A.}~\bibnamefont
  {Buchleitner}},\ and\ \bibinfo {author} {\bibfnamefont {E.~G.}\ \bibnamefont
  {Carnio}},\ }\bibfield  {title} {\bibinfo {title} {An energy estimation
  benchmark for quantum computing hardware},\ }\href
  {https://doi.org/10.1142/S1230161224500069} {\bibfield  {journal} {\bibinfo
  {journal} {Open Systems \& Information Dynamics}\ }\textbf {\bibinfo {volume}
  {31}},\ \bibinfo {pages} {2450006} (\bibinfo {year} {2024})}\BibitemShut
  {NoStop}%
\bibitem [{\citenamefont {Sarovar}\ \emph {et~al.}(2020)\citenamefont
  {Sarovar}, \citenamefont {Proctor}, \citenamefont {Rudinger}, \citenamefont
  {Young}, \citenamefont {Nielsen},\ and\ \citenamefont
  {{Blume-Kohout}}}]{sarovar2020detecting}%
  \BibitemOpen
  \bibfield  {author} {\bibinfo {author} {\bibfnamefont {M.}~\bibnamefont
  {Sarovar}}, \bibinfo {author} {\bibfnamefont {T.}~\bibnamefont {Proctor}},
  \bibinfo {author} {\bibfnamefont {K.}~\bibnamefont {Rudinger}}, \bibinfo
  {author} {\bibfnamefont {K.}~\bibnamefont {Young}}, \bibinfo {author}
  {\bibfnamefont {E.}~\bibnamefont {Nielsen}},\ and\ \bibinfo {author}
  {\bibfnamefont {R.}~\bibnamefont {{Blume-Kohout}}},\ }\bibfield  {title}
  {\bibinfo {title} {Detecting crosstalk errors in quantum information
  processors},\ }\href {https://doi.org/10.22331/q-2020-09-11-321} {\bibfield
  {journal} {\bibinfo  {journal} {Quantum}\ }\textbf {\bibinfo {volume} {4}},\
  \bibinfo {pages} {321} (\bibinfo {year} {2020})}\BibitemShut {NoStop}%
\bibitem [{\citenamefont {Seo}\ \emph {et~al.}(2021)\citenamefont {Seo},
  \citenamefont {Seong},\ and\ \citenamefont
  {Bae}}]{seoMitigationCrosstalkErrors2021}%
  \BibitemOpen
  \bibfield  {author} {\bibinfo {author} {\bibfnamefont {S.}~\bibnamefont
  {Seo}}, \bibinfo {author} {\bibfnamefont {J.}~\bibnamefont {Seong}},\ and\
  \bibinfo {author} {\bibfnamefont {J.}~\bibnamefont {Bae}},\ }\href@noop {}
  {\bibinfo {title} {Mitigation of {{Crosstalk Errors}} in a {{Quantum
  Measurement}} and {{Its Applications}}}} (\bibinfo {year} {2021}),\ \Eprint
  {https://arxiv.org/abs/2112.10651} {arXiv:2112.10651 [quant-ph]} \BibitemShut
  {NoStop}%
\bibitem [{\citenamefont {Ketterer}\ and\ \citenamefont
  {Wellens}(2023)}]{Ketterer2023}%
  \BibitemOpen
  \bibfield  {author} {\bibinfo {author} {\bibfnamefont {A.}~\bibnamefont
  {Ketterer}}\ and\ \bibinfo {author} {\bibfnamefont {T.}~\bibnamefont
  {Wellens}},\ }\bibfield  {title} {\bibinfo {title} {Characterizing crosstalk
  of superconducting transmon processors},\ }\href
  {https://doi.org/10.1103/physrevapplied.20.034065} {\bibfield  {journal}
  {\bibinfo  {journal} {Physical Review Applied}\ }\textbf {\bibinfo {volume}
  {20}},\ \bibinfo {pages} {034065} (\bibinfo {year} {2023})}\BibitemShut
  {NoStop}%
\bibitem [{\citenamefont {Coffman}\ \emph {et~al.}(2000)\citenamefont
  {Coffman}, \citenamefont {Kundu},\ and\ \citenamefont
  {Wootters}}]{coffman2000distributedentanglement}%
  \BibitemOpen
  \bibfield  {author} {\bibinfo {author} {\bibfnamefont {V.}~\bibnamefont
  {Coffman}}, \bibinfo {author} {\bibfnamefont {J.}~\bibnamefont {Kundu}},\
  and\ \bibinfo {author} {\bibfnamefont {W.~K.}\ \bibnamefont {Wootters}},\
  }\bibfield  {title} {\bibinfo {title} {Distributed entanglement},\ }\href
  {https://doi.org/10.1103/PhysRevA.61.052306} {\bibfield  {journal} {\bibinfo
  {journal} {Physical Review A}\ }\textbf {\bibinfo {volume} {61}},\ \bibinfo
  {pages} {052306} (\bibinfo {year} {2000})}\BibitemShut {NoStop}%
\bibitem [{\citenamefont {Werner}(1989)}]{Werner1989mixedstates}%
  \BibitemOpen
  \bibfield  {author} {\bibinfo {author} {\bibfnamefont {R.~F.}\ \bibnamefont
  {Werner}},\ }\bibfield  {title} {\bibinfo {title} {Quantum states with
  einstein-podolsky-rosen correlations admitting a hidden-variable model},\
  }\href {https://doi.org/10.1103/PhysRevA.40.4277} {\bibfield  {journal}
  {\bibinfo  {journal} {Physical Review A}\ }\textbf {\bibinfo {volume} {40}},\
  \bibinfo {pages} {4277} (\bibinfo {year} {1989})}\BibitemShut {NoStop}%
\bibitem [{\citenamefont {Uhlmann}(2000)}]{Uhlmann2000convexroof}%
  \BibitemOpen
  \bibfield  {author} {\bibinfo {author} {\bibfnamefont {A.}~\bibnamefont
  {Uhlmann}},\ }\bibfield  {title} {\bibinfo {title} {Fidelity and concurrence
  of conjugated states},\ }\href {https://doi.org/10.1103/PhysRevA.62.032307}
  {\bibfield  {journal} {\bibinfo  {journal} {Physical Review A}\ }\textbf
  {\bibinfo {volume} {62}},\ \bibinfo {pages} {032307} (\bibinfo {year}
  {2000})}\BibitemShut {NoStop}%
\bibitem [{\citenamefont {Mintert}\ \emph {et~al.}(2005)\citenamefont
  {Mintert}, \citenamefont {Carvalho}, \citenamefont {Kuś},\ and\
  \citenamefont {Buchleitner}}]{mintert2005measures}%
  \BibitemOpen
  \bibfield  {author} {\bibinfo {author} {\bibfnamefont {F.}~\bibnamefont
  {Mintert}}, \bibinfo {author} {\bibfnamefont {A.~R.~R.}\ \bibnamefont
  {Carvalho}}, \bibinfo {author} {\bibfnamefont {M.}~\bibnamefont {Kuś}},\
  and\ \bibinfo {author} {\bibfnamefont {A.}~\bibnamefont {Buchleitner}},\
  }\bibfield  {title} {\bibinfo {title} {Measures and dynamics of entangled
  states},\ }\href
  {https://doi.org/https://doi.org/10.1016/j.physrep.2005.04.006} {\bibfield
  {journal} {\bibinfo  {journal} {Physics Reports}\ }\textbf {\bibinfo {volume}
  {415}},\ \bibinfo {pages} {207} (\bibinfo {year} {2005})}\BibitemShut
  {NoStop}%
\bibitem [{\citenamefont {Aolita}\ and\ \citenamefont
  {Mintert}(2006)}]{aolita2006singleobservable}%
  \BibitemOpen
  \bibfield  {author} {\bibinfo {author} {\bibfnamefont {L.}~\bibnamefont
  {Aolita}}\ and\ \bibinfo {author} {\bibfnamefont {F.}~\bibnamefont
  {Mintert}},\ }\bibfield  {title} {\bibinfo {title} {Measuring multipartite
  concurrence with a single factorizable observable},\ }\href
  {https://doi.org/10.1103/PhysRevLett.97.050501} {\bibfield  {journal}
  {\bibinfo  {journal} {Physical Review Letters}\ }\textbf {\bibinfo {volume}
  {97}},\ \bibinfo {pages} {050501} (\bibinfo {year} {2006})}\BibitemShut
  {NoStop}%
\bibitem [{\citenamefont {Mintert}\ and\ \citenamefont
  {Buchleitner}(2007)}]{mintert2007observablemixed}%
  \BibitemOpen
  \bibfield  {author} {\bibinfo {author} {\bibfnamefont {F.}~\bibnamefont
  {Mintert}}\ and\ \bibinfo {author} {\bibfnamefont {A.}~\bibnamefont
  {Buchleitner}},\ }\bibfield  {title} {\bibinfo {title} {Observable
  entanglement measure for mixed quantum states},\ }\href
  {https://doi.org/10.1103/PhysRevLett.98.140505} {\bibfield  {journal}
  {\bibinfo  {journal} {Physical Review Letters}\ }\textbf {\bibinfo {volume}
  {98}},\ \bibinfo {pages} {140505} (\bibinfo {year} {2007})}\BibitemShut
  {NoStop}%
\bibitem [{\citenamefont {Aolita}\ \emph {et~al.}(2008)\citenamefont {Aolita},
  \citenamefont {Buchleitner},\ and\ \citenamefont
  {Mintert}}]{Aolita2008scalable}%
  \BibitemOpen
  \bibfield  {author} {\bibinfo {author} {\bibfnamefont {L.}~\bibnamefont
  {Aolita}}, \bibinfo {author} {\bibfnamefont {A.}~\bibnamefont
  {Buchleitner}},\ and\ \bibinfo {author} {\bibfnamefont {F.}~\bibnamefont
  {Mintert}},\ }\bibfield  {title} {\bibinfo {title} {Scalable method to
  estimate experimentally the entanglement of multipartite systems},\ }\href
  {https://doi.org/10.1103/PhysRevA.78.022308} {\bibfield  {journal} {\bibinfo
  {journal} {Physical Review A}\ }\textbf {\bibinfo {volume} {78}},\ \bibinfo
  {pages} {022308} (\bibinfo {year} {2008})}\BibitemShut {NoStop}%
\bibitem [{\citenamefont {G{\"u}hne}\ \emph {et~al.}(2007)\citenamefont
  {G{\"u}hne}, \citenamefont {Lu}, \citenamefont {Gao},\ and\ \citenamefont
  {Pan}}]{guhne2007toolbox}%
  \BibitemOpen
  \bibfield  {author} {\bibinfo {author} {\bibfnamefont {O.}~\bibnamefont
  {G{\"u}hne}}, \bibinfo {author} {\bibfnamefont {C.-Y.}\ \bibnamefont {Lu}},
  \bibinfo {author} {\bibfnamefont {W.-B.}\ \bibnamefont {Gao}},\ and\ \bibinfo
  {author} {\bibfnamefont {J.-W.}\ \bibnamefont {Pan}},\ }\bibfield  {title}
  {\bibinfo {title} {Toolbox for entanglement detection and fidelity
  estimation},\ }\href {https://doi.org/10.1103/PhysRevA.76.030305} {\bibfield
  {journal} {\bibinfo  {journal} {Physical Review A}\ }\textbf {\bibinfo
  {volume} {76}},\ \bibinfo {pages} {030305} (\bibinfo {year}
  {2007})}\BibitemShut {NoStop}%
\bibitem [{\citenamefont {Gühne}\ and\ \citenamefont
  {Tóth}(2009)}]{guhne2009entanglement}%
  \BibitemOpen
  \bibfield  {author} {\bibinfo {author} {\bibfnamefont {O.}~\bibnamefont
  {Gühne}}\ and\ \bibinfo {author} {\bibfnamefont {G.}~\bibnamefont {Tóth}},\
  }\bibfield  {title} {\bibinfo {title} {Entanglement detection},\ }\href
  {https://doi.org/10.1016/j.physrep.2009.02.004} {\bibfield  {journal}
  {\bibinfo  {journal} {Physics Reports}\ }\textbf {\bibinfo {volume} {474}},\
  \bibinfo {pages} {1} (\bibinfo {year} {2009})}\BibitemShut {NoStop}%
\bibitem [{\citenamefont {Ketterer}\ \emph {et~al.}(2019)\citenamefont
  {Ketterer}, \citenamefont {Wyderka},\ and\ \citenamefont
  {G{\"u}hne}}]{ketterer2019characterizing}%
  \BibitemOpen
  \bibfield  {author} {\bibinfo {author} {\bibfnamefont {A.}~\bibnamefont
  {Ketterer}}, \bibinfo {author} {\bibfnamefont {N.}~\bibnamefont {Wyderka}},\
  and\ \bibinfo {author} {\bibfnamefont {O.}~\bibnamefont {G{\"u}hne}},\
  }\bibfield  {title} {\bibinfo {title} {Characterizing multipartite
  entanglement with moments of random correlations},\ }\href
  {https://doi.org/10.1103/PhysRevLett.122.120505} {\bibfield  {journal}
  {\bibinfo  {journal} {Physical Review Letters}\ }\textbf {\bibinfo {volume}
  {122}},\ \bibinfo {pages} {120505} (\bibinfo {year} {2019})}\BibitemShut
  {NoStop}%
\bibitem [{\citenamefont {Cramer}\ \emph {et~al.}(2010)\citenamefont {Cramer},
  \citenamefont {Plenio}, \citenamefont {Flammia}, \citenamefont {Somma},
  \citenamefont {Gross}, \citenamefont {Bartlett}, \citenamefont
  {Landon-Cardinal}, \citenamefont {Poulin},\ and\ \citenamefont
  {Liu}}]{cramer2010efficient}%
  \BibitemOpen
  \bibfield  {author} {\bibinfo {author} {\bibfnamefont {M.}~\bibnamefont
  {Cramer}}, \bibinfo {author} {\bibfnamefont {M.~B.}\ \bibnamefont {Plenio}},
  \bibinfo {author} {\bibfnamefont {S.~T.}\ \bibnamefont {Flammia}}, \bibinfo
  {author} {\bibfnamefont {R.}~\bibnamefont {Somma}}, \bibinfo {author}
  {\bibfnamefont {D.}~\bibnamefont {Gross}}, \bibinfo {author} {\bibfnamefont
  {S.~D.}\ \bibnamefont {Bartlett}}, \bibinfo {author} {\bibfnamefont
  {O.}~\bibnamefont {Landon-Cardinal}}, \bibinfo {author} {\bibfnamefont
  {D.}~\bibnamefont {Poulin}},\ and\ \bibinfo {author} {\bibfnamefont {Y.-K.}\
  \bibnamefont {Liu}},\ }\bibfield  {title} {\bibinfo {title} {Efficient
  quantum state tomography},\ }\href {https://doi.org/10.1038/ncomms1147}
  {\bibfield  {journal} {\bibinfo  {journal} {Nature Communications}\ }\textbf
  {\bibinfo {volume} {1}},\ \bibinfo {pages} {149} (\bibinfo {year}
  {2010})}\BibitemShut {NoStop}%
\bibitem [{\citenamefont {Gross}\ \emph {et~al.}(2010)\citenamefont {Gross},
  \citenamefont {Liu}, \citenamefont {Flammia}, \citenamefont {Becker},\ and\
  \citenamefont {Eisert}}]{gross2010quantum}%
  \BibitemOpen
  \bibfield  {author} {\bibinfo {author} {\bibfnamefont {D.}~\bibnamefont
  {Gross}}, \bibinfo {author} {\bibfnamefont {Y.-K.}\ \bibnamefont {Liu}},
  \bibinfo {author} {\bibfnamefont {S.~T.}\ \bibnamefont {Flammia}}, \bibinfo
  {author} {\bibfnamefont {S.}~\bibnamefont {Becker}},\ and\ \bibinfo {author}
  {\bibfnamefont {J.}~\bibnamefont {Eisert}},\ }\bibfield  {title} {\bibinfo
  {title} {Quantum state tomography via compressed sensing},\ }\href
  {https://doi.org/10.1103/PhysRevLett.105.150401} {\bibfield  {journal}
  {\bibinfo  {journal} {Physical Review Letters}\ }\textbf {\bibinfo {volume}
  {105}},\ \bibinfo {pages} {150401} (\bibinfo {year} {2010})}\BibitemShut
  {NoStop}%
\bibitem [{\citenamefont {Torlai}\ \emph {et~al.}(2018)\citenamefont {Torlai},
  \citenamefont {Mazzola}, \citenamefont {Carrasquilla}, \citenamefont
  {Troyer}, \citenamefont {Melko},\ and\ \citenamefont
  {Carleo}}]{torlai2018neural}%
  \BibitemOpen
  \bibfield  {author} {\bibinfo {author} {\bibfnamefont {G.}~\bibnamefont
  {Torlai}}, \bibinfo {author} {\bibfnamefont {G.}~\bibnamefont {Mazzola}},
  \bibinfo {author} {\bibfnamefont {J.}~\bibnamefont {Carrasquilla}}, \bibinfo
  {author} {\bibfnamefont {M.}~\bibnamefont {Troyer}}, \bibinfo {author}
  {\bibfnamefont {R.}~\bibnamefont {Melko}},\ and\ \bibinfo {author}
  {\bibfnamefont {G.}~\bibnamefont {Carleo}},\ }\bibfield  {title} {\bibinfo
  {title} {Neural-network quantum state tomography},\ }\href
  {https://doi.org/10.1038/s41567-018-0048-5} {\bibfield  {journal} {\bibinfo
  {journal} {Nature Physics}\ }\textbf {\bibinfo {volume} {14}},\ \bibinfo
  {pages} {447} (\bibinfo {year} {2018})}\BibitemShut {NoStop}%
\bibitem [{\citenamefont {Lu}\ \emph {et~al.}(2018)\citenamefont {Lu},
  \citenamefont {Zhao}, \citenamefont {Li}, \citenamefont {Yin}, \citenamefont
  {Yuan}, \citenamefont {Hung}, \citenamefont {Chen}, \citenamefont {Li},
  \citenamefont {Liu}, \citenamefont {Peng}, \citenamefont {Liang},
  \citenamefont {Ma}, \citenamefont {Chen},\ and\ \citenamefont
  {Pan}}]{lu_entanglement_2018}%
  \BibitemOpen
  \bibfield  {author} {\bibinfo {author} {\bibfnamefont {H.}~\bibnamefont
  {Lu}}, \bibinfo {author} {\bibfnamefont {Q.}~\bibnamefont {Zhao}}, \bibinfo
  {author} {\bibfnamefont {Z.-D.}\ \bibnamefont {Li}}, \bibinfo {author}
  {\bibfnamefont {X.-F.}\ \bibnamefont {Yin}}, \bibinfo {author} {\bibfnamefont
  {X.}~\bibnamefont {Yuan}}, \bibinfo {author} {\bibfnamefont {J.-C.}\
  \bibnamefont {Hung}}, \bibinfo {author} {\bibfnamefont {L.-K.}\ \bibnamefont
  {Chen}}, \bibinfo {author} {\bibfnamefont {L.}~\bibnamefont {Li}}, \bibinfo
  {author} {\bibfnamefont {N.-L.}\ \bibnamefont {Liu}}, \bibinfo {author}
  {\bibfnamefont {C.-Z.}\ \bibnamefont {Peng}}, \bibinfo {author}
  {\bibfnamefont {Y.-C.}\ \bibnamefont {Liang}}, \bibinfo {author}
  {\bibfnamefont {X.}~\bibnamefont {Ma}}, \bibinfo {author} {\bibfnamefont
  {Y.-A.}\ \bibnamefont {Chen}},\ and\ \bibinfo {author} {\bibfnamefont
  {J.-W.}\ \bibnamefont {Pan}},\ }\bibfield  {title} {\bibinfo {title}
  {Entanglement {Structure}: {Entanglement} {Partitioning} in {Multipartite}
  {Systems} and {Its} {Experimental} {Detection} {Using} {Optimizable}
  {Witnesses}},\ }\href {https://doi.org/10.1103/PhysRevX.8.021072} {\bibfield
  {journal} {\bibinfo  {journal} {Physical Review X}\ }\textbf {\bibinfo
  {volume} {8}},\ \bibinfo {pages} {021072} (\bibinfo {year}
  {2018})}\BibitemShut {NoStop}%
\bibitem [{\citenamefont {Koutný}\ \emph {et~al.}(2023)\citenamefont
  {Koutný}, \citenamefont {Ginés}, \citenamefont {Moczała-Dusanowska},
  \citenamefont {Höfling}, \citenamefont {Schneider}, \citenamefont
  {Predojević},\ and\ \citenamefont {Ježek}}]{koutny2023deeplearning}%
  \BibitemOpen
  \bibfield  {author} {\bibinfo {author} {\bibfnamefont {D.}~\bibnamefont
  {Koutný}}, \bibinfo {author} {\bibfnamefont {L.}~\bibnamefont {Ginés}},
  \bibinfo {author} {\bibfnamefont {M.}~\bibnamefont {Moczała-Dusanowska}},
  \bibinfo {author} {\bibfnamefont {S.}~\bibnamefont {Höfling}}, \bibinfo
  {author} {\bibfnamefont {C.}~\bibnamefont {Schneider}}, \bibinfo {author}
  {\bibfnamefont {A.}~\bibnamefont {Predojević}},\ and\ \bibinfo {author}
  {\bibfnamefont {M.}~\bibnamefont {Ježek}},\ }\bibfield  {title} {\bibinfo
  {title} {Deep learning of quantum entanglement from incomplete
  measurements},\ }\href {https://doi.org/10.1126/sciadv.add7131} {\bibfield
  {journal} {\bibinfo  {journal} {Science Advances}\ }\textbf {\bibinfo
  {volume} {9}},\ \bibinfo {pages} {eadd7131} (\bibinfo {year}
  {2023})}\BibitemShut {NoStop}%
\bibitem [{\citenamefont {Ohnemus}\ \emph {et~al.}(2023)\citenamefont
  {Ohnemus}, \citenamefont {Breuer},\ and\ \citenamefont
  {Ketterer}}]{Ohnemus2023mpc}%
  \BibitemOpen
  \bibfield  {author} {\bibinfo {author} {\bibfnamefont {S.}~\bibnamefont
  {Ohnemus}}, \bibinfo {author} {\bibfnamefont {H.-P.}\ \bibnamefont
  {Breuer}},\ and\ \bibinfo {author} {\bibfnamefont {A.}~\bibnamefont
  {Ketterer}},\ }\bibfield  {title} {\bibinfo {title} {Quantifying
  multiparticle entanglement with randomized measurements},\ }\href
  {https://doi.org/10.1103/PhysRevA.107.042406} {\bibfield  {journal} {\bibinfo
   {journal} {Physical Review A}\ }\textbf {\bibinfo {volume} {107}},\ \bibinfo
  {pages} {042406} (\bibinfo {year} {2023})}\BibitemShut {NoStop}%
\bibitem [{\citenamefont {{IBM Quantum}}()}]{falcon}%
  \BibitemOpen
  \bibfield  {author} {\bibinfo {author} {\bibnamefont {{IBM Quantum}}},\
  }\href {https://docs.quantum.ibm.com/run/processor-types} {\bibinfo {title}
  {{IBM} quantum processor types}},\ \bibinfo {note} {accessed on
  13.05.2024}\BibitemShut {NoStop}%
\bibitem [{\citenamefont {Brunner}\ \emph {et~al.}()\citenamefont {Brunner},
  \citenamefont {Xie}, \citenamefont {Dufour},\ and\ \citenamefont
  {Buchleitner}}]{brunner_inprep}%
  \BibitemOpen
  \bibfield  {author} {\bibinfo {author} {\bibfnamefont {E.}~\bibnamefont
  {Brunner}}, \bibinfo {author} {\bibfnamefont {A.}~\bibnamefont {Xie}},
  \bibinfo {author} {\bibfnamefont {G.}~\bibnamefont {Dufour}},\ and\ \bibinfo
  {author} {\bibfnamefont {A.}~\bibnamefont {Buchleitner}},\ }\href@noop {}
  {\bibinfo {title} {Data-driven approach to mixed-state multipartite
  entanglement characterisation}},\ \Eprint {https://arxiv.org/abs/2407.18014}
  {arXiv:2407.18014} \BibitemShut {NoStop}%
\bibitem [{\citenamefont {Brunner}(2023)}]{brunner_thesis}%
  \BibitemOpen
  \bibfield  {author} {\bibinfo {author} {\bibfnamefont {E.}~\bibnamefont
  {Brunner}},\ }\emph {\bibinfo {title} {Interference and Interactions: Robust
  Signatures of Coherence and Correlations in Many-Body Quantum Systems}},\
  \href {https://doi.org/10.6094/UNIFR/238244} {\bibinfo {type}
  {Dissertation}},\ \bibinfo  {school} {Albert-Ludwigs-Universität Freiburg}
  (\bibinfo {year} {2023})\BibitemShut {NoStop}%
\bibitem [{\citenamefont {McInnes}\ \emph {et~al.}(2018)\citenamefont
  {McInnes}, \citenamefont {Healy},\ and\ \citenamefont
  {Melville}}]{mcinnes2018umap}%
  \BibitemOpen
  \bibfield  {author} {\bibinfo {author} {\bibfnamefont {L.}~\bibnamefont
  {McInnes}}, \bibinfo {author} {\bibfnamefont {J.}~\bibnamefont {Healy}},\
  and\ \bibinfo {author} {\bibfnamefont {J.}~\bibnamefont {Melville}},\
  }\href@noop {} {\bibinfo {title} {Umap: Uniform manifold approximation and
  projection for dimension reduction}} (\bibinfo {year} {2018}),\ \Eprint
  {https://arxiv.org/abs/1802.03426} {arXiv:1802.03426} \BibitemShut {NoStop}%
\bibitem [{\citenamefont {Bengtsson}\ and\ \citenamefont
  {{\.Z}yczkowski}(2017)}]{bengtsson2017geometry}%
  \BibitemOpen
  \bibfield  {author} {\bibinfo {author} {\bibfnamefont {I.}~\bibnamefont
  {Bengtsson}}\ and\ \bibinfo {author} {\bibfnamefont {K.}~\bibnamefont
  {{\.Z}yczkowski}},\ }\href@noop {} {\emph {\bibinfo {title} {Geometry of
  quantum states: an introduction to quantum entanglement}}}\ (\bibinfo
  {publisher} {Cambridge University Press},\ \bibinfo {year}
  {2017})\BibitemShut {NoStop}%
\bibitem [{\citenamefont {Dür}\ \emph {et~al.}(2000)\citenamefont {Dür},
  \citenamefont {Vidal},\ and\ \citenamefont {Cirac}}]{dur2000three}%
  \BibitemOpen
  \bibfield  {author} {\bibinfo {author} {\bibfnamefont {W.}~\bibnamefont
  {Dür}}, \bibinfo {author} {\bibfnamefont {G.}~\bibnamefont {Vidal}},\ and\
  \bibinfo {author} {\bibfnamefont {J.~I.}\ \bibnamefont {Cirac}},\ }\bibfield
  {title} {\bibinfo {title} {Three qubits can be entangled in two inequivalent
  ways},\ }\href {https://doi.org/10.1103/PhysRevA.62.062314} {\bibfield
  {journal} {\bibinfo  {journal} {Physical Review A}\ }\textbf {\bibinfo
  {volume} {62}},\ \bibinfo {pages} {062314} (\bibinfo {year}
  {2000})}\BibitemShut {NoStop}%
\bibitem [{\citenamefont {Greenberger}\ \emph {et~al.}(1990)\citenamefont
  {Greenberger}, \citenamefont {Horne}, \citenamefont {Shimony},\ and\
  \citenamefont {Zeilinger}}]{greenberger1990bell}%
  \BibitemOpen
  \bibfield  {author} {\bibinfo {author} {\bibfnamefont {D.~M.}\ \bibnamefont
  {Greenberger}}, \bibinfo {author} {\bibfnamefont {M.~A.}\ \bibnamefont
  {Horne}}, \bibinfo {author} {\bibfnamefont {A.}~\bibnamefont {Shimony}},\
  and\ \bibinfo {author} {\bibfnamefont {A.}~\bibnamefont {Zeilinger}},\
  }\bibfield  {title} {\bibinfo {title} {Bell’s theorem without
  inequalities},\ }\href@noop {} {\bibfield  {journal} {\bibinfo  {journal}
  {American Journal of Physics}\ }\textbf {\bibinfo {volume} {58}},\ \bibinfo
  {pages} {1131} (\bibinfo {year} {1990})}\BibitemShut {NoStop}%
\bibitem [{\citenamefont {Diker}(2016)}]{diker2016deterministic}%
  \BibitemOpen
  \bibfield  {author} {\bibinfo {author} {\bibfnamefont {F.}~\bibnamefont
  {Diker}},\ }\href@noop {} {\bibinfo {title} {Deterministic construction of
  arbitrary {$ W $} states with quadratically increasing number of two-qubit
  gates}} (\bibinfo {year} {2016}),\ \Eprint {https://arxiv.org/abs/1606.09290}
  {arXiv:1606.09290} \BibitemShut {NoStop}%
\bibitem [{\citenamefont {Woitzik}\ \emph {et~al.}(2020)\citenamefont
  {Woitzik}, \citenamefont {Barkoutsos}, \citenamefont {Wudarski},
  \citenamefont {Buchleitner},\ and\ \citenamefont
  {Tavernelli}}]{woitzikEntanglementProductionConvergence2020}%
  \BibitemOpen
  \bibfield  {author} {\bibinfo {author} {\bibfnamefont {A.~J.~C.}\
  \bibnamefont {Woitzik}}, \bibinfo {author} {\bibfnamefont {P.~K.}\
  \bibnamefont {Barkoutsos}}, \bibinfo {author} {\bibfnamefont
  {F.}~\bibnamefont {Wudarski}}, \bibinfo {author} {\bibfnamefont
  {A.}~\bibnamefont {Buchleitner}},\ and\ \bibinfo {author} {\bibfnamefont
  {I.}~\bibnamefont {Tavernelli}},\ }\bibfield  {title} {\bibinfo {title}
  {Entanglement production and convergence properties of the variational
  quantum eigensolver},\ }\href {https://doi.org/10.1103/PhysRevA.102.042402}
  {\bibfield  {journal} {\bibinfo  {journal} {Physical Review A}\ }\textbf
  {\bibinfo {volume} {102}},\ \bibinfo {pages} {042402} (\bibinfo {year}
  {2020})}\BibitemShut {NoStop}%
\bibitem [{\citenamefont {de~Leon}\ \emph {et~al.}(2021)\citenamefont
  {de~Leon}, \citenamefont {Itoh}, \citenamefont {Kim}, \citenamefont {Mehta},
  \citenamefont {Northup} \emph {et~al.}}]{de2021materials}%
  \BibitemOpen
  \bibfield  {author} {\bibinfo {author} {\bibfnamefont {N.~P.}\ \bibnamefont
  {de~Leon}}, \bibinfo {author} {\bibfnamefont {K.~M.}\ \bibnamefont {Itoh}},
  \bibinfo {author} {\bibfnamefont {D.}~\bibnamefont {Kim}}, \bibinfo {author}
  {\bibfnamefont {K.~K.}\ \bibnamefont {Mehta}}, \bibinfo {author}
  {\bibfnamefont {T.~E.}\ \bibnamefont {Northup}}, \emph {et~al.},\ }\bibfield
  {title} {\bibinfo {title} {Materials challenges and opportunities for quantum
  computing hardware},\ }\href {https://doi.org/10.1126/science.abb2823}
  {\bibfield  {journal} {\bibinfo  {journal} {Science}\ }\textbf {\bibinfo
  {volume} {372}},\ \bibinfo {pages} {eabb2823} (\bibinfo {year}
  {2021})}\BibitemShut {NoStop}%
\bibitem [{\citenamefont {Kjaergaard}\ \emph {et~al.}(2020)\citenamefont
  {Kjaergaard}, \citenamefont {Schwartz}, \citenamefont {Braumüller},
  \citenamefont {Krantz}, \citenamefont {Wang} \emph
  {et~al.}}]{kjaergaard2020superconducting}%
  \BibitemOpen
  \bibfield  {author} {\bibinfo {author} {\bibfnamefont {M.}~\bibnamefont
  {Kjaergaard}}, \bibinfo {author} {\bibfnamefont {M.~E.}\ \bibnamefont
  {Schwartz}}, \bibinfo {author} {\bibfnamefont {J.}~\bibnamefont
  {Braumüller}}, \bibinfo {author} {\bibfnamefont {P.}~\bibnamefont {Krantz}},
  \bibinfo {author} {\bibfnamefont {J.~I.-J.}\ \bibnamefont {Wang}}, \emph
  {et~al.},\ }\bibfield  {title} {\bibinfo {title} {Superconducting qubits:
  Current state of play},\ }\href
  {https://doi.org/10.1146/annurev-conmatphys-031119-050605} {\bibfield
  {journal} {\bibinfo  {journal} {Annual Review of Condensed Matter Physics}\
  }\textbf {\bibinfo {volume} {11}},\ \bibinfo {pages} {369} (\bibinfo {year}
  {2020})}\BibitemShut {NoStop}%
\bibitem [{\citenamefont {Blatt}\ and\ \citenamefont
  {Roos}(2012)}]{blatt2012quantum}%
  \BibitemOpen
  \bibfield  {author} {\bibinfo {author} {\bibfnamefont {R.}~\bibnamefont
  {Blatt}}\ and\ \bibinfo {author} {\bibfnamefont {C.~F.}\ \bibnamefont
  {Roos}},\ }\bibfield  {title} {\bibinfo {title} {Quantum simulations with
  trapped ions},\ }\href {https://doi.org/10.1038/nphys2252} {\bibfield
  {journal} {\bibinfo  {journal} {Nature Physics}\ }\textbf {\bibinfo {volume}
  {8}},\ \bibinfo {pages} {277} (\bibinfo {year} {2012})}\BibitemShut {NoStop}%
\bibitem [{\citenamefont {Bruzewicz}\ \emph {et~al.}(2019)\citenamefont
  {Bruzewicz}, \citenamefont {Chiaverini}, \citenamefont {McConnell},\ and\
  \citenamefont {Sage}}]{bruzewicz2019trapped}%
  \BibitemOpen
  \bibfield  {author} {\bibinfo {author} {\bibfnamefont {C.~D.}\ \bibnamefont
  {Bruzewicz}}, \bibinfo {author} {\bibfnamefont {J.}~\bibnamefont
  {Chiaverini}}, \bibinfo {author} {\bibfnamefont {R.}~\bibnamefont
  {McConnell}},\ and\ \bibinfo {author} {\bibfnamefont {J.~M.}\ \bibnamefont
  {Sage}},\ }\bibfield  {title} {\bibinfo {title} {Trapped-ion quantum
  computing: Progress and challenges},\ }\href
  {https://doi.org/10.1063/1.5088164} {\bibfield  {journal} {\bibinfo
  {journal} {Applied Physics Reviews}\ }\textbf {\bibinfo {volume} {6}},\
  \bibinfo {pages} {021314} (\bibinfo {year} {2019})}\BibitemShut {NoStop}%
\bibitem [{\citenamefont {Nation}\ and\ \citenamefont
  {Treinish}(2023)}]{nation2023suppressing}%
  \BibitemOpen
  \bibfield  {author} {\bibinfo {author} {\bibfnamefont {P.~D.}\ \bibnamefont
  {Nation}}\ and\ \bibinfo {author} {\bibfnamefont {M.}~\bibnamefont
  {Treinish}},\ }\bibfield  {title} {\bibinfo {title} {Suppressing quantum
  circuit errors due to system variability},\ }\href@noop {} {\bibfield
  {journal} {\bibinfo  {journal} {PRX Quantum}\ }\textbf {\bibinfo {volume}
  {4}},\ \bibinfo {pages} {010327} (\bibinfo {year} {2023})}\BibitemShut
  {NoStop}%
\bibitem [{\citenamefont {Cai}\ \emph {et~al.}(2022)\citenamefont {Cai},
  \citenamefont {Babbush}, \citenamefont {Benjamin}, \citenamefont {Endo},
  \citenamefont {Huggins} \emph {et~al.}}]{cai2022quantum}%
  \BibitemOpen
  \bibfield  {author} {\bibinfo {author} {\bibfnamefont {Z.}~\bibnamefont
  {Cai}}, \bibinfo {author} {\bibfnamefont {R.}~\bibnamefont {Babbush}},
  \bibinfo {author} {\bibfnamefont {S.~C.}\ \bibnamefont {Benjamin}}, \bibinfo
  {author} {\bibfnamefont {S.}~\bibnamefont {Endo}}, \bibinfo {author}
  {\bibfnamefont {W.~J.}\ \bibnamefont {Huggins}}, \emph {et~al.},\ }\href@noop
  {} {\bibinfo {title} {Quantum error mitigation}} (\bibinfo {year} {2022}),\
  \Eprint {https://arxiv.org/abs/2210.00921} {arXiv:2210.00921} \BibitemShut
  {NoStop}%
\bibitem [{Note1()}]{Note1}%
  \BibitemOpen
  \bibinfo {note} {This machine belongs to the `Falcon' family of variant r5.11
  \cite {falcon}}\BibitemShut {NoStop}%
\bibitem [{\citenamefont {Treinish}\ \emph {et~al.}(2023)\citenamefont
  {Treinish}, \citenamefont {Gambetta}, \citenamefont {{Soolu Thomas}},
  \citenamefont {Nation}, \citenamefont {{Qiskit-Bot}}, \citenamefont
  {Kassebaum}, \citenamefont {Rodríguez}, \citenamefont {De~La
  Puente~González}, \citenamefont {Lishman}, \citenamefont {{Shaohan Hu}},
  \citenamefont {Bello}, \citenamefont {Arellano}, \citenamefont {Garrison},
  \citenamefont {{Junye Huang}}, \citenamefont {Krsulich}, \citenamefont {Yu},
  \citenamefont {Gacon}, \citenamefont {Marques}, \citenamefont {McKay},
  \citenamefont {Gomez}, \citenamefont {Capelluto}, \citenamefont
  {{Travis-S-IBM}}, \citenamefont {Mitchell}, \citenamefont {Panigrahi},
  \citenamefont {Hartman}, \citenamefont {{Lerongil}}, \citenamefont {{Rafey
  Iqbal Rahman}}, \citenamefont {Wood}, \citenamefont {{Toshinari Itoko}},\
  and\ \citenamefont {Pozas-Kerstjens}}]{qiskit2021}%
  \BibitemOpen
  \bibfield  {author} {\bibinfo {author} {\bibfnamefont {M.}~\bibnamefont
  {Treinish}}, \bibinfo {author} {\bibfnamefont {J.}~\bibnamefont {Gambetta}},
  \bibinfo {author} {\bibnamefont {{Soolu Thomas}}}, \bibinfo {author}
  {\bibfnamefont {P.}~\bibnamefont {Nation}}, \bibinfo {author} {\bibnamefont
  {{Qiskit-Bot}}}, \bibinfo {author} {\bibfnamefont {P.}~\bibnamefont
  {Kassebaum}}, \bibinfo {author} {\bibfnamefont {D.~M.}\ \bibnamefont
  {Rodríguez}}, \bibinfo {author} {\bibfnamefont {S.}~\bibnamefont {De~La
  Puente~González}}, \bibinfo {author} {\bibfnamefont {J.}~\bibnamefont
  {Lishman}}, \bibinfo {author} {\bibnamefont {{Shaohan Hu}}}, \bibinfo
  {author} {\bibfnamefont {L.}~\bibnamefont {Bello}}, \bibinfo {author}
  {\bibfnamefont {E.}~\bibnamefont {Arellano}}, \bibinfo {author}
  {\bibfnamefont {J.}~\bibnamefont {Garrison}}, \bibinfo {author} {\bibnamefont
  {{Junye Huang}}}, \bibinfo {author} {\bibfnamefont {K.}~\bibnamefont
  {Krsulich}}, \bibinfo {author} {\bibfnamefont {J.}~\bibnamefont {Yu}},
  \bibinfo {author} {\bibfnamefont {J.}~\bibnamefont {Gacon}}, \bibinfo
  {author} {\bibfnamefont {M.}~\bibnamefont {Marques}}, \bibinfo {author}
  {\bibfnamefont {D.}~\bibnamefont {McKay}}, \bibinfo {author} {\bibfnamefont
  {J.}~\bibnamefont {Gomez}}, \bibinfo {author} {\bibfnamefont
  {L.}~\bibnamefont {Capelluto}}, \bibinfo {author} {\bibnamefont
  {{Travis-S-IBM}}}, \bibinfo {author} {\bibfnamefont {A.}~\bibnamefont
  {Mitchell}}, \bibinfo {author} {\bibfnamefont {A.}~\bibnamefont {Panigrahi}},
  \bibinfo {author} {\bibfnamefont {K.}~\bibnamefont {Hartman}}, \bibinfo
  {author} {\bibnamefont {{Lerongil}}}, \bibinfo {author} {\bibnamefont {{Rafey
  Iqbal Rahman}}}, \bibinfo {author} {\bibfnamefont {S.}~\bibnamefont {Wood}},
  \bibinfo {author} {\bibnamefont {{Toshinari Itoko}}},\ and\ \bibinfo {author}
  {\bibfnamefont {A.}~\bibnamefont {Pozas-Kerstjens}},\ }\href
  {https://doi.org/10.5281/ZENODO.7897504} {\bibinfo {title}
  {Qiskit/qiskit-metapackage: Qiskit 0.43.0}} (\bibinfo {year}
  {2023})\BibitemShut {NoStop}%
\bibitem [{\citenamefont {Terhal}(2002)}]{TERHAL2002313}%
  \BibitemOpen
  \bibfield  {author} {\bibinfo {author} {\bibfnamefont {B.~M.}\ \bibnamefont
  {Terhal}},\ }\bibfield  {title} {\bibinfo {title} {Detecting quantum
  entanglement},\ }\href@noop {} {\bibfield  {journal} {\bibinfo  {journal}
  {Theoretical Computer Science}\ }\textbf {\bibinfo {volume} {287}},\ \bibinfo
  {pages} {313} (\bibinfo {year} {2002})}\BibitemShut {NoStop}%
\bibitem [{\citenamefont {Horodecki}\ \emph {et~al.}(2009)\citenamefont
  {Horodecki}, \citenamefont {Horodecki}, \citenamefont {Horodecki},\ and\
  \citenamefont {Horodecki}}]{RevModPhys.81.865}%
  \BibitemOpen
  \bibfield  {author} {\bibinfo {author} {\bibfnamefont {R.}~\bibnamefont
  {Horodecki}}, \bibinfo {author} {\bibfnamefont {P.}~\bibnamefont
  {Horodecki}}, \bibinfo {author} {\bibfnamefont {M.}~\bibnamefont
  {Horodecki}},\ and\ \bibinfo {author} {\bibfnamefont {K.}~\bibnamefont
  {Horodecki}},\ }\bibfield  {title} {\bibinfo {title} {Quantum entanglement},\
  }\href {https://doi.org/10.1103/RevModPhys.81.865} {\bibfield  {journal}
  {\bibinfo  {journal} {Rev. Mod. Phys.}\ }\textbf {\bibinfo {volume} {81}},\
  \bibinfo {pages} {865} (\bibinfo {year} {2009})}\BibitemShut {NoStop}%
\bibitem [{\citenamefont {G{\"u}hne}\ and\ \citenamefont
  {T{\'o}th}(2009)}]{GUHNE20091}%
  \BibitemOpen
  \bibfield  {author} {\bibinfo {author} {\bibfnamefont {O.}~\bibnamefont
  {G{\"u}hne}}\ and\ \bibinfo {author} {\bibfnamefont {G.}~\bibnamefont
  {T{\'o}th}},\ }\bibfield  {title} {\bibinfo {title} {Entanglement
  detection},\ }\href@noop {} {\bibfield  {journal} {\bibinfo  {journal}
  {Physics Reports}\ }\textbf {\bibinfo {volume} {474}},\ \bibinfo {pages} {1}
  (\bibinfo {year} {2009})}\BibitemShut {NoStop}%
\bibitem [{\citenamefont {Chruściński}\ and\ \citenamefont
  {Sarbicki}(2014)}]{Chruscinski2014}%
  \BibitemOpen
  \bibfield  {author} {\bibinfo {author} {\bibfnamefont {D.}~\bibnamefont
  {Chruściński}}\ and\ \bibinfo {author} {\bibfnamefont {G.}~\bibnamefont
  {Sarbicki}},\ }\bibfield  {title} {\bibinfo {title} {Entanglement witnesses:
  construction, analysis and classification},\ }\href
  {https://doi.org/10.1088/1751-8113/47/48/483001} {\bibfield  {journal}
  {\bibinfo  {journal} {Journal of Physics A: Mathematical and Theoretical}\
  }\textbf {\bibinfo {volume} {47}},\ \bibinfo {pages} {483001} (\bibinfo
  {year} {2014})}\BibitemShut {NoStop}%
\bibitem [{\citenamefont {Friis}\ \emph {et~al.}(2019)\citenamefont {Friis},
  \citenamefont {Vitagliano}, \citenamefont {Malik},\ and\ \citenamefont
  {Huber}}]{Huber2019}%
  \BibitemOpen
  \bibfield  {author} {\bibinfo {author} {\bibfnamefont {N.}~\bibnamefont
  {Friis}}, \bibinfo {author} {\bibfnamefont {G.}~\bibnamefont {Vitagliano}},
  \bibinfo {author} {\bibfnamefont {M.}~\bibnamefont {Malik}},\ and\ \bibinfo
  {author} {\bibfnamefont {M.}~\bibnamefont {Huber}},\ }\bibfield  {title}
  {\bibinfo {title} {Entanglement certification from theory to experiment},\
  }\href@noop {} {\bibfield  {journal} {\bibinfo  {journal} {Nature Reviews
  Physics}\ }\textbf {\bibinfo {volume} {1}},\ \bibinfo {pages} {72} (\bibinfo
  {year} {2019})}\BibitemShut {NoStop}%
\bibitem [{\citenamefont {T\'oth}\ and\ \citenamefont
  {G\"uhne}(2005{\natexlab{a}})}]{PhysRevA.72.022340}%
  \BibitemOpen
  \bibfield  {author} {\bibinfo {author} {\bibfnamefont {G.}~\bibnamefont
  {T\'oth}}\ and\ \bibinfo {author} {\bibfnamefont {O.}~\bibnamefont
  {G\"uhne}},\ }\bibfield  {title} {\bibinfo {title} {Entanglement detection in
  the stabilizer formalism},\ }\href
  {https://doi.org/10.1103/PhysRevA.72.022340} {\bibfield  {journal} {\bibinfo
  {journal} {Physical Review A}\ }\textbf {\bibinfo {volume} {72}},\ \bibinfo
  {pages} {022340} (\bibinfo {year} {2005}{\natexlab{a}})}\BibitemShut
  {NoStop}%
\bibitem [{\citenamefont {T\'oth}\ and\ \citenamefont
  {G\"uhne}(2005{\natexlab{b}})}]{PhysRevLett.94.060501}%
  \BibitemOpen
  \bibfield  {author} {\bibinfo {author} {\bibfnamefont {G.}~\bibnamefont
  {T\'oth}}\ and\ \bibinfo {author} {\bibfnamefont {O.}~\bibnamefont
  {G\"uhne}},\ }\bibfield  {title} {\bibinfo {title} {Detecting genuine
  multipartite entanglement with two local measurements},\ }\href
  {https://doi.org/10.1103/PhysRevLett.94.060501} {\bibfield  {journal}
  {\bibinfo  {journal} {Physical Review Letters}\ }\textbf {\bibinfo {volume}
  {94}},\ \bibinfo {pages} {060501} (\bibinfo {year}
  {2005}{\natexlab{b}})}\BibitemShut {NoStop}%
\bibitem [{Note2()}]{Note2}%
  \BibitemOpen
  \bibinfo {note} {It is worth mentioning that \protect \texttt {ibm\protect
  \_ehningen}, \protect \texttt {ibm\protect \_cairo}, \protect \texttt
  {ibmq\protect \_ehningen} and \protect \texttt {ibm\protect \_hanoi} are of
  the same processor type, i.e., Falcon r5.11, whereas \protect \texttt
  {ibmq\protect \_mumbai} is of type Falcon r5.10 \cite
  {pelofske2022quantumvolume}. In this latter reference `r5.10' is mislabeled
  as `r5.1'.}\BibitemShut {Stop}%
\bibitem [{ibm()}]{ibm_NoiseModel}%
  \BibitemOpen
  \href
  {https://qiskit.github.io/qiskit-aer/stubs/qiskit_aer.noise.NoiseModel.html#qiskit_aer.noise.NoiseModel.from_backend}
  {\bibinfo {title} {{Qiskit Aer 0.13.1 API Documentation - NoiseModel}}},\
  \bibinfo {note} {accessed on 13.05.2024}\BibitemShut {NoStop}%
\bibitem [{\citenamefont {Życzkowski}\ \emph {et~al.}(1998)\citenamefont
  {Życzkowski}, \citenamefont {Horodecki}, \citenamefont {Sanpera},\ and\
  \citenamefont {Lewenstein}}]{zyczkowski_volume_1998}%
  \BibitemOpen
  \bibfield  {author} {\bibinfo {author} {\bibfnamefont {K.}~\bibnamefont
  {Życzkowski}}, \bibinfo {author} {\bibfnamefont {P.}~\bibnamefont
  {Horodecki}}, \bibinfo {author} {\bibfnamefont {A.}~\bibnamefont {Sanpera}},\
  and\ \bibinfo {author} {\bibfnamefont {M.}~\bibnamefont {Lewenstein}},\
  }\bibfield  {title} {\bibinfo {title} {Volume of the set of separable
  states},\ }\href {https://doi.org/10.1103/PhysRevA.58.883} {\bibfield
  {journal} {\bibinfo  {journal} {Physical Review A}\ }\textbf {\bibinfo
  {volume} {58}},\ \bibinfo {pages} {883} (\bibinfo {year} {1998})}\BibitemShut
  {NoStop}%
\bibitem [{\citenamefont {Fine}\ \emph {et~al.}(2005)\citenamefont {Fine},
  \citenamefont {Mintert},\ and\ \citenamefont
  {Buchleitner}}]{fine_equilibrium_2005}%
  \BibitemOpen
  \bibfield  {author} {\bibinfo {author} {\bibfnamefont {B.~V.}\ \bibnamefont
  {Fine}}, \bibinfo {author} {\bibfnamefont {F.}~\bibnamefont {Mintert}},\ and\
  \bibinfo {author} {\bibfnamefont {A.}~\bibnamefont {Buchleitner}},\
  }\bibfield  {title} {\bibinfo {title} {Equilibrium entanglement vanishes at
  finite temperature},\ }\href {https://doi.org/10.1103/PhysRevB.71.153105}
  {\bibfield  {journal} {\bibinfo  {journal} {Physical Review B}\ }\textbf
  {\bibinfo {volume} {71}},\ \bibinfo {pages} {153105} (\bibinfo {year}
  {2005})}\BibitemShut {NoStop}%
\bibitem [{Note3()}]{Note3}%
  \BibitemOpen
  \bibinfo {note} {This is done by generating a random Haar unitary in
  $2^{N_i}$ dimensions (with $N_i$ the number of qubits in part $P_i$) and
  applying it to an arbitrary but fixed $N_i$-qubit state. For this we use the
  QuTiP \cite {johansson_qutip_2013} implementation of the algorithm proposed
  in \cite {mezzadri_how_2007}.}\BibitemShut {Stop}%
\bibitem [{\citenamefont {Walschaers}\ \emph {et~al.}(2016)\citenamefont
  {Walschaers}, \citenamefont {Kuipers}, \citenamefont {Urbina}, \citenamefont
  {Mayer}, \citenamefont {Tichy}, \citenamefont {Richter},\ and\ \citenamefont
  {Buchleitner}}]{walschaers_statistical_2016}%
  \BibitemOpen
  \bibfield  {author} {\bibinfo {author} {\bibfnamefont {M.}~\bibnamefont
  {Walschaers}}, \bibinfo {author} {\bibfnamefont {J.}~\bibnamefont {Kuipers}},
  \bibinfo {author} {\bibfnamefont {J.-D.}\ \bibnamefont {Urbina}}, \bibinfo
  {author} {\bibfnamefont {K.}~\bibnamefont {Mayer}}, \bibinfo {author}
  {\bibfnamefont {M.~C.}\ \bibnamefont {Tichy}}, \bibinfo {author}
  {\bibfnamefont {K.}~\bibnamefont {Richter}},\ and\ \bibinfo {author}
  {\bibfnamefont {A.}~\bibnamefont {Buchleitner}},\ }\bibfield  {title}
  {\bibinfo {title} {Statistical benchmark for {BosonSampling}},\ }\href
  {https://doi.org/10.1088/1367-2630/18/3/032001} {\bibfield  {journal}
  {\bibinfo  {journal} {New Journal of Physics}\ }\textbf {\bibinfo {volume}
  {18}},\ \bibinfo {pages} {032001} (\bibinfo {year} {2016})}\BibitemShut
  {NoStop}%
\bibitem [{Note4()}]{Note4}%
  \BibitemOpen
  \bibinfo {note} {Recall that the test dataset was generated in conjunction
  with, and therefore carries the same statistical properties as, the training
  dataset. The accuracy score for other datasets will be, in general,
  different.}\BibitemShut {Stop}%
\bibitem [{Note5()}]{Note5}%
  \BibitemOpen
  \bibinfo {note} {For partitions with more than one part (all except for `6'),
  we have tested both adjacent and distant sets of qubits for the different
  parts. However, this distinction did not lead to any difference in the
  results, which we therefore do not differentiate in Figs.~\ref
  {fig:entanglement_partition_assignment}.}\BibitemShut {Stop}%
\bibitem [{Note6()}]{Note6}%
  \BibitemOpen
  \bibinfo {note} {Note that by using the `ladder' circuit we could
  occasionally (for individual realizations but not on average over the
  randomly selected sets of connected qubits) also witness GHZ-type
  entanglement up to six qubits, which is not the case for the `star' circuit
  results shown in Fig.~\ref {fig:witness_scaling}.}\BibitemShut {Stop}%
\bibitem [{\citenamefont {\ifmmode~\dot{Z}\else
  \.{Z}\fi{}yczkowski}(1999)}]{Zyczkowski1999volume}%
  \BibitemOpen
  \bibfield  {author} {\bibinfo {author} {\bibfnamefont {K.}~\bibnamefont
  {\ifmmode~\dot{Z}\else \.{Z}\fi{}yczkowski}},\ }\bibfield  {title} {\bibinfo
  {title} {Volume of the set of separable states. ii},\ }\href
  {https://doi.org/10.1103/PhysRevA.60.3496} {\bibfield  {journal} {\bibinfo
  {journal} {Physical Review A}\ }\textbf {\bibinfo {volume} {60}},\ \bibinfo
  {pages} {3496} (\bibinfo {year} {1999})}\BibitemShut {NoStop}%
\bibitem [{\citenamefont {Bae}\ \emph {et~al.}(2009)\citenamefont {Bae},
  \citenamefont {Tiersch}, \citenamefont {Sauer}, \citenamefont {de~Melo},
  \citenamefont {Mintert}, \citenamefont {Hiesmayr},\ and\ \citenamefont
  {Buchleitner}}]{Bae_bound_2009}%
  \BibitemOpen
  \bibfield  {author} {\bibinfo {author} {\bibfnamefont {J.}~\bibnamefont
  {Bae}}, \bibinfo {author} {\bibfnamefont {M.}~\bibnamefont {Tiersch}},
  \bibinfo {author} {\bibfnamefont {S.}~\bibnamefont {Sauer}}, \bibinfo
  {author} {\bibfnamefont {F.}~\bibnamefont {de~Melo}}, \bibinfo {author}
  {\bibfnamefont {F.}~\bibnamefont {Mintert}}, \bibinfo {author} {\bibfnamefont
  {B.}~\bibnamefont {Hiesmayr}},\ and\ \bibinfo {author} {\bibfnamefont
  {A.}~\bibnamefont {Buchleitner}},\ }\bibfield  {title} {\bibinfo {title}
  {Detection and typicality of bound entangled states},\ }\href
  {https://doi.org/10.1103/PhysRevA.80.022317} {\bibfield  {journal} {\bibinfo
  {journal} {Phys. Rev. A}\ }\textbf {\bibinfo {volume} {80}},\ \bibinfo
  {pages} {022317} (\bibinfo {year} {2009})}\BibitemShut {NoStop}%
\bibitem [{\citenamefont {Hiesmayr}(2021)}]{hiesmayr_free_2021}%
  \BibitemOpen
  \bibfield  {author} {\bibinfo {author} {\bibfnamefont {B.~C.}\ \bibnamefont
  {Hiesmayr}},\ }\bibfield  {title} {\bibinfo {title} {Free versus bound
  entanglement, a {NP}-hard problem tackled by machine learning},\ }\href
  {https://doi.org/10.1038/s41598-021-98523-6} {\bibfield  {journal} {\bibinfo
  {journal} {Sci Rep}\ }\textbf {\bibinfo {volume} {11}},\ \bibinfo {pages}
  {19739} (\bibinfo {year} {2021})}\BibitemShut {NoStop}%
\bibitem [{\citenamefont {Schollw{\"o}ck}(2011)}]{schollwock2011density}%
  \BibitemOpen
  \bibfield  {author} {\bibinfo {author} {\bibfnamefont {U.}~\bibnamefont
  {Schollw{\"o}ck}},\ }\bibfield  {title} {\bibinfo {title} {The density-matrix
  renormalization group in the age of matrix product states},\ }\href@noop {}
  {\bibfield  {journal} {\bibinfo  {journal} {Annals of physics}\ }\textbf
  {\bibinfo {volume} {326}},\ \bibinfo {pages} {96} (\bibinfo {year}
  {2011})}\BibitemShut {NoStop}%
\bibitem [{\citenamefont {Cirac}\ \emph {et~al.}(2021)\citenamefont {Cirac},
  \citenamefont {Perez-Garcia}, \citenamefont {Schuch},\ and\ \citenamefont
  {Verstraete}}]{cirac2021matrix}%
  \BibitemOpen
  \bibfield  {author} {\bibinfo {author} {\bibfnamefont {J.~I.}\ \bibnamefont
  {Cirac}}, \bibinfo {author} {\bibfnamefont {D.}~\bibnamefont {Perez-Garcia}},
  \bibinfo {author} {\bibfnamefont {N.}~\bibnamefont {Schuch}},\ and\ \bibinfo
  {author} {\bibfnamefont {F.}~\bibnamefont {Verstraete}},\ }\bibfield  {title}
  {\bibinfo {title} {Matrix product states and projected entangled pair states:
  Concepts, symmetries, theorems},\ }\href@noop {} {\bibfield  {journal}
  {\bibinfo  {journal} {Reviews of Modern Physics}\ }\textbf {\bibinfo {volume}
  {93}},\ \bibinfo {pages} {045003} (\bibinfo {year} {2021})}\BibitemShut
  {NoStop}%
\bibitem [{\citenamefont {Gorbachev}\ and\ \citenamefont
  {Trubilko}(2000)}]{Gorbachev2000}%
  \BibitemOpen
  \bibfield  {author} {\bibinfo {author} {\bibfnamefont {V.~N.}\ \bibnamefont
  {Gorbachev}}\ and\ \bibinfo {author} {\bibfnamefont {A.~I.}\ \bibnamefont
  {Trubilko}},\ }\bibfield  {title} {\bibinfo {title} {Quantum teleportation of
  an einstein-podolsy-rosen pair using an entangled three-particle state},\
  }\href {https://doi.org/10.1134/1.1334979} {\bibfield  {journal} {\bibinfo
  {journal} {Journal of Experimental and Theoretical Physics}\ }\textbf
  {\bibinfo {volume} {91}},\ \bibinfo {pages} {894} (\bibinfo {year}
  {2000})}\BibitemShut {NoStop}%
\bibitem [{\citenamefont {Zhang}\ \emph {et~al.}(2022)\citenamefont {Zhang},
  \citenamefont {Srinivasan}, \citenamefont {Sundaresan}, \citenamefont
  {Bogorin}, \citenamefont {Martin}, \citenamefont {Hertzberg}, \citenamefont
  {Timmerwilke}, \citenamefont {Pritchett}, \citenamefont {Yau}, \citenamefont
  {Wang}, \citenamefont {Landers}, \citenamefont {Lewandowski}, \citenamefont
  {Narasgond}, \citenamefont {Rosenblatt}, \citenamefont {Keefe}, \citenamefont
  {Lauer}, \citenamefont {Rothwell}, \citenamefont {McClure}, \citenamefont
  {Dial}, \citenamefont {Orcutt}, \citenamefont {Brink},\ and\ \citenamefont
  {Chow}}]{zhang2022highperformance}%
  \BibitemOpen
  \bibfield  {author} {\bibinfo {author} {\bibfnamefont {E.~J.}\ \bibnamefont
  {Zhang}}, \bibinfo {author} {\bibfnamefont {S.}~\bibnamefont {Srinivasan}},
  \bibinfo {author} {\bibfnamefont {N.}~\bibnamefont {Sundaresan}}, \bibinfo
  {author} {\bibfnamefont {D.~F.}\ \bibnamefont {Bogorin}}, \bibinfo {author}
  {\bibfnamefont {Y.}~\bibnamefont {Martin}}, \bibinfo {author} {\bibfnamefont
  {J.~B.}\ \bibnamefont {Hertzberg}}, \bibinfo {author} {\bibfnamefont
  {J.}~\bibnamefont {Timmerwilke}}, \bibinfo {author} {\bibfnamefont {E.~J.}\
  \bibnamefont {Pritchett}}, \bibinfo {author} {\bibfnamefont {J.-B.}\
  \bibnamefont {Yau}}, \bibinfo {author} {\bibfnamefont {C.}~\bibnamefont
  {Wang}}, \bibinfo {author} {\bibfnamefont {W.}~\bibnamefont {Landers}},
  \bibinfo {author} {\bibfnamefont {E.~P.}\ \bibnamefont {Lewandowski}},
  \bibinfo {author} {\bibfnamefont {A.}~\bibnamefont {Narasgond}}, \bibinfo
  {author} {\bibfnamefont {S.}~\bibnamefont {Rosenblatt}}, \bibinfo {author}
  {\bibfnamefont {G.~A.}\ \bibnamefont {Keefe}}, \bibinfo {author}
  {\bibfnamefont {I.}~\bibnamefont {Lauer}}, \bibinfo {author} {\bibfnamefont
  {M.~B.}\ \bibnamefont {Rothwell}}, \bibinfo {author} {\bibfnamefont {D.~T.}\
  \bibnamefont {McClure}}, \bibinfo {author} {\bibfnamefont {O.~E.}\
  \bibnamefont {Dial}}, \bibinfo {author} {\bibfnamefont {J.~S.}\ \bibnamefont
  {Orcutt}}, \bibinfo {author} {\bibfnamefont {M.}~\bibnamefont {Brink}},\ and\
  \bibinfo {author} {\bibfnamefont {J.~M.}\ \bibnamefont {Chow}},\ }\bibfield
  {title} {\bibinfo {title} {High-performance superconducting quantum
  processors via laser annealing of transmon qubits},\ }\href
  {https://doi.org/10.1126/sciadv.abi6690} {\bibfield  {journal} {\bibinfo
  {journal} {Science Advances}\ }\textbf {\bibinfo {volume} {8}},\ \bibinfo
  {pages} {eabi6690} (\bibinfo {year} {2022})}\BibitemShut {NoStop}%
\bibitem [{\citenamefont {Vidal}\ and\ \citenamefont
  {Werner}(2002)}]{vidal_computable_2002}%
  \BibitemOpen
  \bibfield  {author} {\bibinfo {author} {\bibfnamefont {G.}~\bibnamefont
  {Vidal}}\ and\ \bibinfo {author} {\bibfnamefont {R.~F.}\ \bibnamefont
  {Werner}},\ }\bibfield  {title} {\bibinfo {title} {Computable measure of
  entanglement},\ }\href {https://doi.org/10.1103/PhysRevA.65.032314}
  {\bibfield  {journal} {\bibinfo  {journal} {Physical Review A}\ }\textbf
  {\bibinfo {volume} {65}},\ \bibinfo {pages} {032314} (\bibinfo {year}
  {2002})}\BibitemShut {NoStop}%
\bibitem [{\citenamefont {Plenio}(2005)}]{plenio_logarithmic_2005}%
  \BibitemOpen
  \bibfield  {author} {\bibinfo {author} {\bibfnamefont {M.~B.}\ \bibnamefont
  {Plenio}},\ }\bibfield  {title} {\bibinfo {title} {Logarithmic {Negativity}:
  {A} {Full} {Entanglement} {Monotone} {That} is not {Convex}},\ }\href
  {https://doi.org/10.1103/PhysRevLett.95.090503} {\bibfield  {journal}
  {\bibinfo  {journal} {Physical Review Letters}\ }\textbf {\bibinfo {volume}
  {95}},\ \bibinfo {pages} {090503} (\bibinfo {year} {2005})}\BibitemShut
  {NoStop}%
\bibitem [{\citenamefont {Peres}(1996)}]{peres_separability_1996}%
  \BibitemOpen
  \bibfield  {author} {\bibinfo {author} {\bibfnamefont {A.}~\bibnamefont
  {Peres}},\ }\bibfield  {title} {\bibinfo {title} {Separability {Criterion}
  for {Density} {Matrices}},\ }\href
  {https://doi.org/10.1103/PhysRevLett.77.1413} {\bibfield  {journal} {\bibinfo
   {journal} {Physical Review Letters}\ }\textbf {\bibinfo {volume} {77}},\
  \bibinfo {pages} {1413} (\bibinfo {year} {1996})}\BibitemShut {NoStop}%
\bibitem [{\citenamefont {Horodecki}\ \emph {et~al.}(1996)\citenamefont
  {Horodecki}, \citenamefont {Horodecki},\ and\ \citenamefont
  {Horodecki}}]{horodecki_separability_1996}%
  \BibitemOpen
  \bibfield  {author} {\bibinfo {author} {\bibfnamefont {M.}~\bibnamefont
  {Horodecki}}, \bibinfo {author} {\bibfnamefont {P.}~\bibnamefont
  {Horodecki}},\ and\ \bibinfo {author} {\bibfnamefont {R.}~\bibnamefont
  {Horodecki}},\ }\bibfield  {title} {\bibinfo {title} {Separability of mixed
  states: necessary and sufficient conditions},\ }\href
  {https://doi.org/10.1016/S0375-9601(96)00706-2} {\bibfield  {journal}
  {\bibinfo  {journal} {Physics Letters A}\ }\textbf {\bibinfo {volume}
  {223}},\ \bibinfo {pages} {1} (\bibinfo {year} {1996})}\BibitemShut {NoStop}%
\bibitem [{\citenamefont {van Enk}\ and\ \citenamefont
  {Beenakker}(2012)}]{van_enk_measuring_2012}%
  \BibitemOpen
  \bibfield  {author} {\bibinfo {author} {\bibfnamefont {S.~J.}\ \bibnamefont
  {van Enk}}\ and\ \bibinfo {author} {\bibfnamefont {C.~W.~J.}\ \bibnamefont
  {Beenakker}},\ }\bibfield  {title} {\bibinfo {title} {Measuring
  $\Tr \rho^n$
  on {Single} {Copies} of $\rho$
  {Using} {Random} {Measurements}},\ }\href
  {https://doi.org/10.1103/PhysRevLett.108.110503} {\bibfield  {journal}
  {\bibinfo  {journal} {Physical Review Letters}\ }\textbf {\bibinfo {volume}
  {108}},\ \bibinfo {pages} {110503} (\bibinfo {year} {2012})}\BibitemShut
  {NoStop}%
\bibitem [{\citenamefont {Elben}\ \emph {et~al.}(2019)\citenamefont {Elben},
  \citenamefont {Vermersch}, \citenamefont {Roos},\ and\ \citenamefont
  {Zoller}}]{elben_statistical_2019}%
  \BibitemOpen
  \bibfield  {author} {\bibinfo {author} {\bibfnamefont {A.}~\bibnamefont
  {Elben}}, \bibinfo {author} {\bibfnamefont {B.}~\bibnamefont {Vermersch}},
  \bibinfo {author} {\bibfnamefont {C.~F.}\ \bibnamefont {Roos}},\ and\
  \bibinfo {author} {\bibfnamefont {P.}~\bibnamefont {Zoller}},\ }\bibfield
  {title} {\bibinfo {title} {Statistical correlations between locally
  randomized measurements: {A} toolbox for probing entanglement in many-body
  quantum states},\ }\href {https://doi.org/10.1103/PhysRevA.99.052323}
  {\bibfield  {journal} {\bibinfo  {journal} {Physical Review A}\ }\textbf
  {\bibinfo {volume} {99}},\ \bibinfo {pages} {052323} (\bibinfo {year}
  {2019})}\BibitemShut {NoStop}%
\bibitem [{\citenamefont {Knips}\ \emph {et~al.}(2020)\citenamefont {Knips},
  \citenamefont {Dziewior}, \citenamefont {Kłobus}, \citenamefont {Laskowski},
  \citenamefont {Paterek}, \citenamefont {Shadbolt}, \citenamefont
  {Weinfurter},\ and\ \citenamefont {Meinecke}}]{knips_multipartite_2020}%
  \BibitemOpen
  \bibfield  {author} {\bibinfo {author} {\bibfnamefont {L.}~\bibnamefont
  {Knips}}, \bibinfo {author} {\bibfnamefont {J.}~\bibnamefont {Dziewior}},
  \bibinfo {author} {\bibfnamefont {W.}~\bibnamefont {Kłobus}}, \bibinfo
  {author} {\bibfnamefont {W.}~\bibnamefont {Laskowski}}, \bibinfo {author}
  {\bibfnamefont {T.}~\bibnamefont {Paterek}}, \bibinfo {author} {\bibfnamefont
  {P.~J.}\ \bibnamefont {Shadbolt}}, \bibinfo {author} {\bibfnamefont
  {H.}~\bibnamefont {Weinfurter}},\ and\ \bibinfo {author} {\bibfnamefont
  {J.~D.~A.}\ \bibnamefont {Meinecke}},\ }\bibfield  {title} {\bibinfo {title}
  {Multipartite entanglement analysis from random correlations},\ }\href
  {https://doi.org/10.1038/s41534-020-0281-5} {\bibfield  {journal} {\bibinfo
  {journal} {npj Quantum Information}\ }\textbf {\bibinfo {volume} {6}},\
  \bibinfo {pages} {1} (\bibinfo {year} {2020})}\BibitemShut {NoStop}%
\bibitem [{\citenamefont {Ketterer}\ \emph {et~al.}(2022)\citenamefont
  {Ketterer}, \citenamefont {Imai}, \citenamefont {Wyderka},\ and\
  \citenamefont {Gühne}}]{ketterer_statistically_2022}%
  \BibitemOpen
  \bibfield  {author} {\bibinfo {author} {\bibfnamefont {A.}~\bibnamefont
  {Ketterer}}, \bibinfo {author} {\bibfnamefont {S.}~\bibnamefont {Imai}},
  \bibinfo {author} {\bibfnamefont {N.}~\bibnamefont {Wyderka}},\ and\ \bibinfo
  {author} {\bibfnamefont {O.}~\bibnamefont {Gühne}},\ }\bibfield  {title}
  {\bibinfo {title} {Statistically significant tests of multiparticle quantum
  correlations based on randomized measurements},\ }\href
  {https://doi.org/10.1103/PhysRevA.106.L010402} {\bibfield  {journal}
  {\bibinfo  {journal} {Physical Review A}\ }\textbf {\bibinfo {volume}
  {106}},\ \bibinfo {pages} {L010402} (\bibinfo {year} {2022})}\BibitemShut
  {NoStop}%
\bibitem [{\citenamefont {Wyderka}\ \emph {et~al.}(2023)\citenamefont
  {Wyderka}, \citenamefont {Ketterer}, \citenamefont {Imai}, \citenamefont
  {B{\"o}nsel}, \citenamefont {Jones}, \citenamefont {Kirby}, \citenamefont
  {Yu},\ and\ \citenamefont {G{\"u}hne}}]{wyderka2023complete}%
  \BibitemOpen
  \bibfield  {author} {\bibinfo {author} {\bibfnamefont {N.}~\bibnamefont
  {Wyderka}}, \bibinfo {author} {\bibfnamefont {A.}~\bibnamefont {Ketterer}},
  \bibinfo {author} {\bibfnamefont {S.}~\bibnamefont {Imai}}, \bibinfo {author}
  {\bibfnamefont {J.~L.}\ \bibnamefont {B{\"o}nsel}}, \bibinfo {author}
  {\bibfnamefont {D.~E.}\ \bibnamefont {Jones}}, \bibinfo {author}
  {\bibfnamefont {B.~T.}\ \bibnamefont {Kirby}}, \bibinfo {author}
  {\bibfnamefont {X.-D.}\ \bibnamefont {Yu}},\ and\ \bibinfo {author}
  {\bibfnamefont {O.}~\bibnamefont {G{\"u}hne}},\ }\bibfield  {title} {\bibinfo
  {title} {Complete characterization of quantum correlations by randomized
  measurements},\ }\href@noop {} {\bibfield  {journal} {\bibinfo  {journal}
  {Physical Review Letters}\ }\textbf {\bibinfo {volume} {131}},\ \bibinfo
  {pages} {090201} (\bibinfo {year} {2023})}\BibitemShut {NoStop}%
\bibitem [{\citenamefont {Brunner}\ \emph {et~al.}(2022)\citenamefont
  {Brunner}, \citenamefont {Buchleitner},\ and\ \citenamefont
  {Dufour}}]{brunner_many-body_2022}%
  \BibitemOpen
  \bibfield  {author} {\bibinfo {author} {\bibfnamefont {E.}~\bibnamefont
  {Brunner}}, \bibinfo {author} {\bibfnamefont {A.}~\bibnamefont
  {Buchleitner}},\ and\ \bibinfo {author} {\bibfnamefont {G.}~\bibnamefont
  {Dufour}},\ }\bibfield  {title} {\bibinfo {title} {Many-body coherence and
  entanglement probed by randomized correlation measurements},\ }\href
  {https://doi.org/10.1103/PhysRevResearch.4.043101} {\bibfield  {journal}
  {\bibinfo  {journal} {Physical Review Research}\ }\textbf {\bibinfo {volume}
  {4}},\ \bibinfo {pages} {043101} (\bibinfo {year} {2022})}\BibitemShut
  {NoStop}%
\bibitem [{\citenamefont {Cao}\ \emph {et~al.}(2019)\citenamefont {Cao},
  \citenamefont {Spielmann}, \citenamefont {Qiu}, \citenamefont {Huang},
  \citenamefont {Ibrahim}, \citenamefont {Hill}, \citenamefont {Zhang},
  \citenamefont {Mundlos}, \citenamefont {Christiansen}, \citenamefont
  {Steemers}, \citenamefont {Trapnell},\ and\ \citenamefont
  {Shendure}}]{cao_single-cell_2019}%
  \BibitemOpen
  \bibfield  {author} {\bibinfo {author} {\bibfnamefont {J.}~\bibnamefont
  {Cao}}, \bibinfo {author} {\bibfnamefont {M.}~\bibnamefont {Spielmann}},
  \bibinfo {author} {\bibfnamefont {X.}~\bibnamefont {Qiu}}, \bibinfo {author}
  {\bibfnamefont {X.}~\bibnamefont {Huang}}, \bibinfo {author} {\bibfnamefont
  {D.~M.}\ \bibnamefont {Ibrahim}}, \bibinfo {author} {\bibfnamefont {A.~J.}\
  \bibnamefont {Hill}}, \bibinfo {author} {\bibfnamefont {F.}~\bibnamefont
  {Zhang}}, \bibinfo {author} {\bibfnamefont {S.}~\bibnamefont {Mundlos}},
  \bibinfo {author} {\bibfnamefont {L.}~\bibnamefont {Christiansen}}, \bibinfo
  {author} {\bibfnamefont {F.~J.}\ \bibnamefont {Steemers}}, \bibinfo {author}
  {\bibfnamefont {C.}~\bibnamefont {Trapnell}},\ and\ \bibinfo {author}
  {\bibfnamefont {J.}~\bibnamefont {Shendure}},\ }\bibfield  {title} {\bibinfo
  {title} {The single-cell transcriptional landscape of mammalian
  organogenesis},\ }\href {https://doi.org/10.1038/s41586-019-0969-x}
  {\bibfield  {journal} {\bibinfo  {journal} {Nature}\ }\textbf {\bibinfo
  {volume} {566}},\ \bibinfo {pages} {496} (\bibinfo {year}
  {2019})}\BibitemShut {NoStop}%
\bibitem [{\citenamefont {Diaz-Papkovich}\ \emph {et~al.}(2019)\citenamefont
  {Diaz-Papkovich}, \citenamefont {Anderson-Trocmé}, \citenamefont
  {Ben-Eghan},\ and\ \citenamefont {Gravel}}]{diaz-papkovich_umap_2019}%
  \BibitemOpen
  \bibfield  {author} {\bibinfo {author} {\bibfnamefont {A.}~\bibnamefont
  {Diaz-Papkovich}}, \bibinfo {author} {\bibfnamefont {L.}~\bibnamefont
  {Anderson-Trocmé}}, \bibinfo {author} {\bibfnamefont {C.}~\bibnamefont
  {Ben-Eghan}},\ and\ \bibinfo {author} {\bibfnamefont {S.}~\bibnamefont
  {Gravel}},\ }\bibfield  {title} {\bibinfo {title} {{UMAP} reveals cryptic
  population structure and phenotype heterogeneity in large genomic cohorts},\
  }\href {https://doi.org/10.1371/journal.pgen.1008432} {\bibfield  {journal}
  {\bibinfo  {journal} {PLOS Genetics}\ }\textbf {\bibinfo {volume} {15}},\
  \bibinfo {pages} {e1008432} (\bibinfo {year} {2019})}\BibitemShut {NoStop}%
\bibitem [{\citenamefont {Carter}\ \emph {et~al.}(2019)\citenamefont {Carter},
  \citenamefont {Armstrong}, \citenamefont {Schubert}, \citenamefont
  {Johnson},\ and\ \citenamefont {Olah}}]{carter_activation_2019}%
  \BibitemOpen
  \bibfield  {author} {\bibinfo {author} {\bibfnamefont {S.}~\bibnamefont
  {Carter}}, \bibinfo {author} {\bibfnamefont {Z.}~\bibnamefont {Armstrong}},
  \bibinfo {author} {\bibfnamefont {L.}~\bibnamefont {Schubert}}, \bibinfo
  {author} {\bibfnamefont {I.}~\bibnamefont {Johnson}},\ and\ \bibinfo {author}
  {\bibfnamefont {C.}~\bibnamefont {Olah}},\ }\bibfield  {title} {\bibinfo
  {title} {Activation {Atlas}},\ }\href
  {https://doi.org/10.23915/distill.00015} {\bibfield  {journal} {\bibinfo
  {journal} {Distill}\ }\textbf {\bibinfo {volume} {4}},\ \bibinfo {pages}
  {e15} (\bibinfo {year} {2019})}\BibitemShut {NoStop}%
\bibitem [{\citenamefont {Ali}\ \emph {et~al.}(2019)\citenamefont {Ali},
  \citenamefont {Jones}, \citenamefont {Xie},\ and\ \citenamefont
  {Williams}}]{ali_timecluster_2019}%
  \BibitemOpen
  \bibfield  {author} {\bibinfo {author} {\bibfnamefont {M.}~\bibnamefont
  {Ali}}, \bibinfo {author} {\bibfnamefont {M.~W.}\ \bibnamefont {Jones}},
  \bibinfo {author} {\bibfnamefont {X.}~\bibnamefont {Xie}},\ and\ \bibinfo
  {author} {\bibfnamefont {M.}~\bibnamefont {Williams}},\ }\bibfield  {title}
  {\bibinfo {title} {{TimeCluster}: dimension reduction applied to temporal
  data for visual analytics},\ }\href
  {https://doi.org/10.1007/s00371-019-01673-y} {\bibfield  {journal} {\bibinfo
  {journal} {The Visual Computer}\ }\textbf {\bibinfo {volume} {35}},\ \bibinfo
  {pages} {1013} (\bibinfo {year} {2019})}\BibitemShut {NoStop}%
\bibitem [{\citenamefont {Kullback}\ and\ \citenamefont
  {Leibler}(1951)}]{kullback_information_1951}%
  \BibitemOpen
  \bibfield  {author} {\bibinfo {author} {\bibfnamefont {S.}~\bibnamefont
  {Kullback}}\ and\ \bibinfo {author} {\bibfnamefont {R.~A.}\ \bibnamefont
  {Leibler}},\ }\bibfield  {title} {\bibinfo {title} {On {Information} and
  {Sufficiency}},\ }\href {https://doi.org/10.1214/aoms/1177729694} {\bibfield
  {journal} {\bibinfo  {journal} {The Annals of Mathematical Statistics}\
  }\textbf {\bibinfo {volume} {22}},\ \bibinfo {pages} {79} (\bibinfo {year}
  {1951})}\BibitemShut {NoStop}%
\bibitem [{\citenamefont {Pedregosa}\ \emph {et~al.}(2011)\citenamefont
  {Pedregosa}, \citenamefont {Varoquaux}, \citenamefont {Gramfort},
  \citenamefont {Michel}, \citenamefont {Thirion}, \citenamefont {Grisel},
  \citenamefont {Blondel}, \citenamefont {Prettenhofer}, \citenamefont {Weiss},
  \citenamefont {Dubourg}, \citenamefont {Vanderplas}, \citenamefont {Passos},
  \citenamefont {Cournapeau}, \citenamefont {Brucher}, \citenamefont {Perrot},\
  and\ \citenamefont {Duchesnay}}]{scikit-learn}%
  \BibitemOpen
  \bibfield  {author} {\bibinfo {author} {\bibfnamefont {F.}~\bibnamefont
  {Pedregosa}}, \bibinfo {author} {\bibfnamefont {G.}~\bibnamefont
  {Varoquaux}}, \bibinfo {author} {\bibfnamefont {A.}~\bibnamefont {Gramfort}},
  \bibinfo {author} {\bibfnamefont {V.}~\bibnamefont {Michel}}, \bibinfo
  {author} {\bibfnamefont {B.}~\bibnamefont {Thirion}}, \bibinfo {author}
  {\bibfnamefont {O.}~\bibnamefont {Grisel}}, \bibinfo {author} {\bibfnamefont
  {M.}~\bibnamefont {Blondel}}, \bibinfo {author} {\bibfnamefont
  {P.}~\bibnamefont {Prettenhofer}}, \bibinfo {author} {\bibfnamefont
  {R.}~\bibnamefont {Weiss}}, \bibinfo {author} {\bibfnamefont
  {V.}~\bibnamefont {Dubourg}}, \bibinfo {author} {\bibfnamefont
  {J.}~\bibnamefont {Vanderplas}}, \bibinfo {author} {\bibfnamefont
  {A.}~\bibnamefont {Passos}}, \bibinfo {author} {\bibfnamefont
  {D.}~\bibnamefont {Cournapeau}}, \bibinfo {author} {\bibfnamefont
  {M.}~\bibnamefont {Brucher}}, \bibinfo {author} {\bibfnamefont
  {M.}~\bibnamefont {Perrot}},\ and\ \bibinfo {author} {\bibfnamefont
  {E.}~\bibnamefont {Duchesnay}},\ }\bibfield  {title} {\bibinfo {title}
  {Scikit-learn: Machine learning in {P}ython},\ }\href@noop {} {\bibfield
  {journal} {\bibinfo  {journal} {Journal of Machine Learning Research}\
  }\textbf {\bibinfo {volume} {12}},\ \bibinfo {pages} {2825} (\bibinfo {year}
  {2011})}\BibitemShut {NoStop}%
\bibitem [{\citenamefont {Pelofske}\ \emph {et~al.}(2022)\citenamefont
  {Pelofske}, \citenamefont {Bärtschi},\ and\ \citenamefont
  {Eidenbenz}}]{pelofske2022quantumvolume}%
  \BibitemOpen
  \bibfield  {author} {\bibinfo {author} {\bibfnamefont {E.}~\bibnamefont
  {Pelofske}}, \bibinfo {author} {\bibfnamefont {A.}~\bibnamefont
  {Bärtschi}},\ and\ \bibinfo {author} {\bibfnamefont {S.}~\bibnamefont
  {Eidenbenz}},\ }\bibfield  {title} {\bibinfo {title} {Quantum volume in
  practice: What users can expect from nisq devices},\ }\href
  {https://doi.org/10.1109/TQE.2022.3184764} {\bibfield  {journal} {\bibinfo
  {journal} {IEEE Transactions on Quantum Engineering}\ }\textbf {\bibinfo
  {volume} {3}},\ \bibinfo {pages} {1} (\bibinfo {year} {2022})}\BibitemShut
  {NoStop}%
\bibitem [{\citenamefont {Johansson}\ \emph {et~al.}(2013)\citenamefont
  {Johansson}, \citenamefont {Nation},\ and\ \citenamefont
  {Nori}}]{johansson_qutip_2013}%
  \BibitemOpen
  \bibfield  {author} {\bibinfo {author} {\bibfnamefont {J.~R.}\ \bibnamefont
  {Johansson}}, \bibinfo {author} {\bibfnamefont {P.~D.}\ \bibnamefont
  {Nation}},\ and\ \bibinfo {author} {\bibfnamefont {F.}~\bibnamefont {Nori}},\
  }\bibfield  {title} {\bibinfo {title} {{QuTiP} 2: {A} {Python} framework for
  the dynamics of open quantum systems},\ }\href
  {https://doi.org/10.1016/j.cpc.2012.11.019} {\bibfield  {journal} {\bibinfo
  {journal} {Computer Physics Communications}\ }\textbf {\bibinfo {volume}
  {184}},\ \bibinfo {pages} {1234} (\bibinfo {year} {2013})}\BibitemShut
  {NoStop}%
\bibitem [{\citenamefont {Mezzadri}(2007)}]{mezzadri_how_2007}%
  \BibitemOpen
  \bibfield  {author} {\bibinfo {author} {\bibfnamefont {F.}~\bibnamefont
  {Mezzadri}},\ }\href@noop {} {\bibinfo {title} {How to generate random
  matrices from the classical compact groups}} (\bibinfo {year} {2007}),\
  \Eprint {https://arxiv.org/abs/math-ph/0609050} {arXiv:math-ph/0609050}
  \BibitemShut {NoStop}%
\end{thebibliography}

%

\end{document}